\documentclass{article}

\usepackage{arxiv}

\usepackage[utf8]{inputenc} % allow utf-8 input
\usepackage[T1]{fontenc}    % use 8-bit T1 fonts
\usepackage{hyperref}       % hyperlinks
\usepackage{url}            % simple URL typesetting
\usepackage{booktabs}       % professional-quality tables
\usepackage{amsfonts}       % blackboard math symbols
\usepackage{nicefrac}       % compact symbols for 1/2, etc.
\usepackage{microtype}      % microtypography
\usepackage{lipsum}		% Can be removed after putting your text content
\usepackage{graphicx}
\usepackage{natbib}
\usepackage{doi}

\usepackage{fancyvrb}
\usepackage{nameref}
\usepackage{color, colortbl}
\usepackage{miller} %for Miller notation
\usepackage{tabularx} %for tables
\usepackage{pifont} %for \ding symbols
\usepackage{hhline}
\usepackage{multirow}
\usepackage{amssymb}
\usepackage{tikz}
\usepackage{ulem}
\usepackage{bm}
\usepackage{verbatim}
\usepackage{caption}
\usepackage{subcaption}
\definecolor{Gray}{gray}{0.9}

\usepackage{upgreek}
\usepackage{import}
\usepackage{adjustbox}
\usepackage{mathtools}
\usepackage{float}
\usepackage{graphicx}
\usepackage{textcomp}
\usepackage{gensymb}
\usepackage{booktabs}
\usepackage{siunitx}
\usepackage{pifont}
\usepackage{amsmath}
\usepackage{hyperref}

\title{Microstructure quality control of steels using deep learning}

\author{ \href{https://orcid.org/0000-0002-0916-5990}{\includegraphics[scale=0.06]{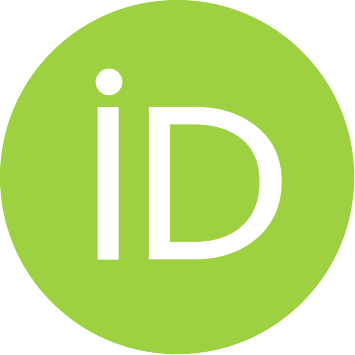}\hspace{1mm}Ali Riza Durmaz} \\
	Group of Meso and Micromechanics\\
	Fraunhofer Institute for Mechanics of Materials IWM\\
	Freiburg im Breisgau, Germany \\
	\texttt{ali.riza.durmaz@iwm.fraunhofer.de} \\
	\And
	Sai Teja Potu \\
        Group of Meso and Micromechanics\\
	Fraunhofer Institute for Mechanics of Materials IWM\\
	Freiburg im Breisgau, Germany \\
        \And
	Daniel Romich \\
	Materials Technology\\
	Schaeffler Technologies AG \& Co. KG \\
        Schweinfurt, Germany \\
        \And
	Johannes Möller \\
	Materials Technology\\
	Schaeffler Technologies AG \& Co. KG \\
        Schweinfurt, Germany \\
        \And
	Ralf Nützel \\
	Materials Technology\\
	Schaeffler Technologies AG \& Co. KG \\
        Schweinfurt, Germany \\
}

\renewcommand{\shorttitle}{Steel microstructure assessment using deep learning}

\hypersetup{
pdftitle={Microstructure quality control of steels using deep learning},
pdfsubject={},
pdfauthor={Ali Riza Durmaz, Sai Teja Potu, Daniel Romich, Johannes Möller, Ralf Nützel},
pdfkeywords={Quality control, Microstructure, Grain size, Steel, Martensite, Bainite, and Deep learning},
}

\begin{document}
\maketitle

\begin{abstract}
	In quality control, microstructures are investigated rigorously to ensure structural integrity, exclude the presence of critical volume defects, and validate the formation of the target microstructure. For quenched, hierarchically-structured steels, the morphology of the bainitic and martensitic microstructures are of major concern to guarantee the reliability of the material under service conditions. Therefore, industries conduct small sample-size inspections of materials cross-sections through metallographers to validate the needle morphology of such microstructures. We demonstrate round-robin test results revealing that this visual grading is afflicted by pronounced subjectivity despite the thorough training of personnel. Instead, we propose a deep learning image classification approach that distinguishes steels based on their microstructure type and classifies their needle length alluding to the ISO 643 grain size assessment standard. This classification approach facilitates the reliable, objective, and automated classification of hierarchically structured steels. Specifically, an accuracy of 96\% and roughly 91\% is attained for the distinction of martensite/bainite subtypes and needle length, respectively. This is achieved on an image dataset that contains significant variance and labeling noise as it is acquired over more than ten years from multiple plants, alloys, etchant applications, and light optical microscopes by many metallographers (raters). Interpretability analysis gives insights into the decision-making of these models and allows for estimating their generalization capability.
\end{abstract}


\section{Introduction}
Materials in many applications are exposed to complicated loading conditions. Along the value chain, components and materials therein are exposed to process scatter at all stages. This establishes the demand for quality control. The presence of major structural defects in components can be excluded through a variety of non-destructive testing methods which exploit ultrasonic or magnetic sensing principles for instance. Most sensor principles that are applicable in-line, however, retrieve integral information of large volumes as they exhibit neither an adequate spatial resolution nor signal sensitivity to measure subtle microstructural heterogeneity. Moreover, most sensors provide compounded information on residual stresses, defect density, grain size, chemical composition, and more. 

\begin{figure}[htbp]
    \centering
    %\footnotesize
    \includegraphics[width=\textwidth]{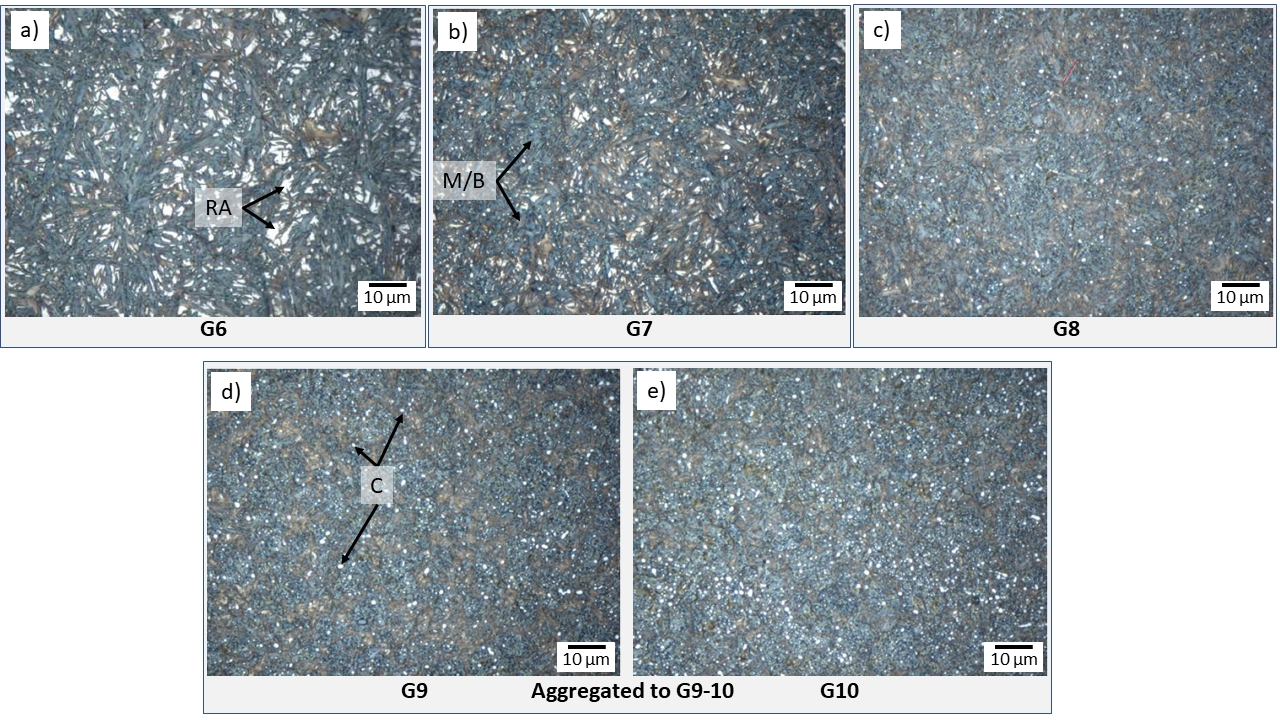} 
    \def\svgwidth{\textwidth}
    \caption{Martensitic steels distinguished by their needle morphology. The arising martensite/bainite (M/B), retained austenite (RA), and carbide (C) constituents are annotated in the figure. The subfigure captions represent the structure codes associated with the grain size according to the ISO 643 standard. Note that this set of images does not represent the variance of the complete dataset as it only takes a single subtype into consideration. A random set of images showcasing a more realistic representation of the data scatter is depicted in Figure \ref{fig:Overview_variance}.}
    \label{fig:Overview_needle_morphologies}
\end{figure}

Therefore, when particular microstructural aspects such as grain size distribution are of interest, most industries fall back on destructive sectioning and direct imaging methodologies, which are being performed on small sample sizes. A typical example of this is hierarchically-structured steel microstructures, such as martensite or bainite, for which the quantification of feature sizes in the primary microstructure is only possible by imaging. Trained metallographers prepare metallographic cross-sections of components through consecutive cutting, polishing, and etching and then image them using light optical microscopy. The resulting micrographs, depending on the component's intended application and loading paths, can be inspected with respect to different microstructural aspects. In bearing steels in which often plate martensite or bainite is present, the contrasted cross-section shows acicular, i.e. needle-shaped, structures (\cite{bepari2017surface}). In this case, the so-called \textit{needle length} affects the material's resistance to rolling contact fatigue (\cite{shur2005physical}) and is thus of pronounced interest. These microstructures are the consequence of the partitioning of the high-temperature austenite phase to a martensitic or bainitic microstructure. Some examples of martensitic steels with varying cooling rates culminating in different morphologies and phase compositions are depicted in Figure \ref{fig:Overview_needle_morphologies}. The microstructures are composed of martensite or bainite (`M/B') as the primary microstructure as well as dispersed minor phases such as retained austenite (bright constituents annotated with `RA') and carbide particles annotated with `C'. While both minor phases occur bright, the carbide particles are either circular or elliptic and slightly smaller than the irregular-shaped retained austenite constituents. Depending on the exact treatment conditions both minor phases are differently distributed.

In this work, two classification problems for hierarchically-structured steel microstructures are tackled. The first is the classification of the needle length as depicted in Figure \ref{fig:Overview_needle_morphologies} and listed in Table \ref{tab:Needle_morphology_classes}. This classification is inspired by the micrographic grain size categorization standard according to ISO 643 which is usually applied for equiaxed and unimodally distributed microstructures. Henceforth, we refer to this task as `grain size', `needle morphology', or `needle length' classification. The classification is applied to bearing steels that underwent different hardening heat treatments. Specifically, through-hardening steels such as 100Cr6, 100CrMnSi6-4 in bainitic (B) and martensitic (G) states as well as quenched and tempered martensitic C56E2 steel (M) are taken into consideration. A distinction between these three subtypes is also of interest as these entail different material properties. This poses the second classification objective. In the following, we refer to the joint information of microstructure subtype and needle length, e.g. `G7', as \textit{structure code}. By attempting the microstructure subtype distinction, we investigate whether nuanced differences between bainitic and martensitic steel variants (\cite{hillert1995nature}) can be identified by computer vision approaches. In contrast to efforts by \cite{muller2020classification, gola2018advanced, zhu2022feature}, we attempt a macroscopic distinction on large field-of-view light optical micrographs without discerning substructures at a prior austenite grain level. Since the bainite and martensite subtype labels in the work at hand are provided from the quantitative temperature-time profile during cooling, this classification problem resembles the one presented by \cite{bulgarevich2019automatic}. The classification of needle length following ISO 643 has not been reported in the literature to the best of our knowledge. Published work at the intersection between DL and metallography, to date, entailed image datasets acquired under comparatively controlled and repeatable conditions, see \cite{decost2019high, durmaz2021deep}. In this work, on the other hand, the difficulty lies in the many degrees of freedom present in the process chain causing a profound data variance.

\begin{table}[htbp]
        \centering
       %\small
       \caption{Steel grade families and grain sizes considered in this study. The classes corresponding to the rows with gray background color are discarded for model training. There are needle length ranges for typical and sporadically occurring large needles associated with the mean and maximum criterion, respectively. Generally, the needle length ranges for both criteria are disjunct. Depending on the measured lengths and their frequency, a metallographer decides which criterion to apply for the micrograph classification. The exact thresholds can not be provided due to confidentiality. }
        \newcolumntype{P}[1]{>{\centering\arraybackslash}p{#1}}
        \begin{tabularx}{\linewidth}{P{0.20\textwidth--2\tabcolsep-1.3333\arrayrulewidth}  P{0.15\textwidth-2\tabcolsep-1.3333\arrayrulewidth} P{0.15\textwidth-2\tabcolsep-1.3333\arrayrulewidth}  P{0.23\textwidth-2\tabcolsep-1.3333\arrayrulewidth}P{0.23\textwidth-2\tabcolsep-1.3333\arrayrulewidth}   
	} 
            \toprule
             \textbf{Material subtype}& \textbf{Grain size (ISO 643)} & \textbf{Structure code} & \textbf{Needle morphology class (mean crit.) }&\textbf{Needle morphology class (max crit.)}\\
            \midrule
            \multirow{4}{*}{Martensitic 100Cr6 (G)} &\cellcolor{Gray} 6 & \cellcolor{Gray} G6    & \cellcolor{Gray} Coarse acicular to acicular     & \cellcolor{Gray} Coarse Acicular \\
             & 7 & G7   & Acicular & Coarse acicular to acicular \\
            & 8 & G8 & Fine acicular & Acicular \\
            & 9 -- 10 & G9--G10 & Fine acicular to structureless & Fine acicular\\
        \midrule
             \multirow{4}{*}{Martensitic C56E2 (M)} & \cellcolor{Gray} 6 & \cellcolor{Gray} M6    & \cellcolor{Gray} Coarse acicular to acicular    & \cellcolor{Gray} Coarse Acicular  \\
           & \cellcolor{Gray} 7 & \cellcolor{Gray} M7   & \cellcolor{Gray} Acicular & \cellcolor{Gray} Coarse acicular to acicular  \\
             & 8 & M8 & Fine acicular & Acicular \\
               & 9 -- 10& M9--M10 & Fine acicular to structureless & Fine acicular\\
        \midrule
           \multirow{4}{*}{Bainitic 100Cr6 (B)}   & \cellcolor{Gray} 6 & \cellcolor{Gray} B6    & \cellcolor{Gray} Coarse acicular to acicular    & \cellcolor{Gray} Coarse Acicular \\
            & 7 & B7   & Acicular & Coarse acicular to acicular \\
            & 8 & B8 & Fine acicular & Acicular \\

               & 9 -- 10  & B9--B10 & Fine acicular to structureless & Fine acicular  \\
        \bottomrule
        \end{tabularx}
        \label{tab:Needle_morphology_classes}
\end{table}

The dataset utilized in this work exhibits a pronounced variance as it covers multiple alloys, heat treatments, polishing protocols, storage times, etchant concentrations, etching durations, and image acquisition parameters. All images are acquired in industrial testing laboratories over a time span of more than 10 years, and their needle length labels (6--10) are assigned by numerous metallographers. A single micrograph is typically inspected by a single metallographer. As the distinction of ISO 643 grain sizes 9 and 10 does not have any application relevance for these bearing steels, both classes in this work are aggregated to the classes 9--10 for each microstructure subtype. Further, since image instances with coarse acicular needles are very scarce in industrial processing, and thus data availability is low, they will be excluded from the classification task, see Table \ref{tab:Needle_morphology_classes}.
Segregation in the material, especially of carbon, chromium, and nickel can influence the local formation of the hardened microstructure and cause fluctuations in the needle morphology. Depending on the heterogeneity of the microstructure, different criteria have been applied to rate the grain size, see Table \ref{tab:Needle_morphology_classes} and Section \ref{sec:data_set_stats}. \par

\begin{figure}[htbp]
    \centering
    %\footnotesize
    \includegraphics[width=\textwidth]{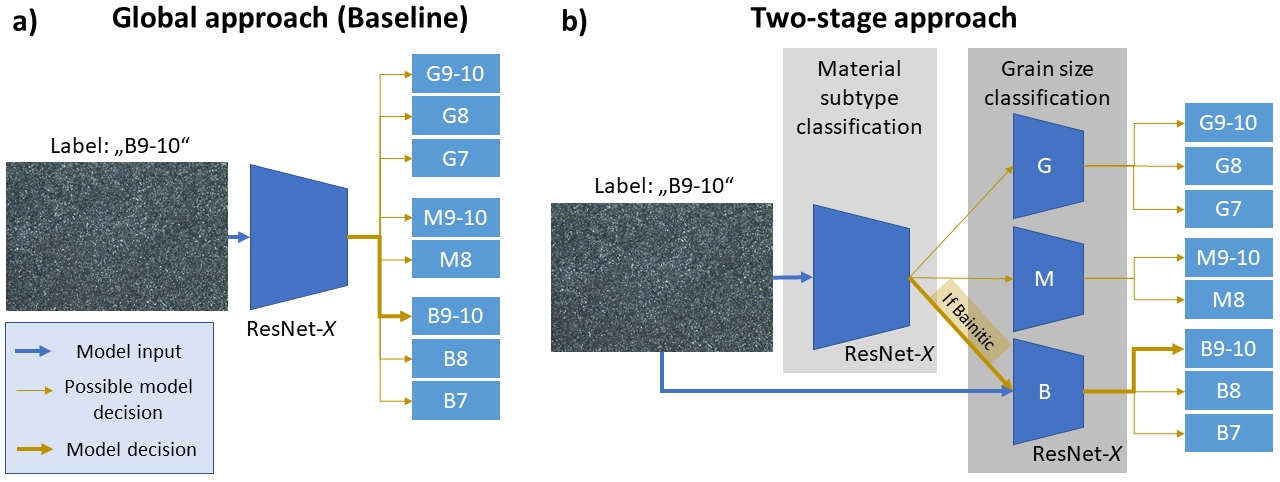} 
    \def\svgwidth{\textwidth}
    %\fontsize{12}{12}\selectfont
    %\input{mpnn_scheme.pdf_tex}
    \caption{A comparison of a single multi-class classification (\textbf{A}) setting with the proposed two-stage approach (\textbf{B}) of first determining the material subtype and then the needle morphology.}
    \label{fig:Classification_approach}
\end{figure}

In an attempt to render both image classification tasks automated and objective, we apply different deep learning (DL) methodologies. To be precise, two approaches are employed which are illustrated in Figure \ref{fig:Classification_approach}a and \ref{fig:Classification_approach}b. The first represents a single multi-class classification model which confronts both classification challenges simultaneously, i.e. direct prediction of the structure code. In contrast, a second methodology is presented in which one model categorizes the micrographs based on the microstructure subtype they are showing, and another model distinguishes the image in terms of its needle morphology. Depending on the first model's subtype prediction, a dedicated needle morphology classifier model is selected to perform the needle length classification. Comparison of both approaches allows investigation of whether decomposing the problem and addressing the partial tasks with specialized models can be beneficial.

In order to evaluate the inter-rater reliability of both image classification tasks, a round-robin test is designed and presented here. Thereby, the subjectivity of the subtype and needle morphology classification can be assessed. Further, the round-robin results act as a baseline for the computational assessment.

\section{Results}

In the following, the overall results are presented for both strategies, the global model and the two-stage (2S) approach. The results are listed in Table \ref{tab:ml_model_performance}. Two common deep learning architectures are taken into consideration, ResNet-50 and ResNet-18. Moreover, the `fw' annotation in the table indicates models trained with frozen weights in the feature extractor portion of the model. As an evaluation metric, accuracy is provided. The accuracy, for the data and models at hand, virtually coincides with the F\textsubscript{1}-score. In a repeatability study, where three distinct test-train splits were sampled for the bainite `B' grain size classification using three-fold cross-validation, the data sampling-induced fluctuations were confirmed to be negligible. Therefore, we report single training performances here.
               
\begin{table}[h]  
    \centering
    \caption{The overall model prediction accuracies achieved on the complete test set. The abbreviation `fw' indicates models which were trained with frozen weights of the feature extractor portion. The specifications `global' and `2S' refer to direct structure code prediction and the two-stage approach depicted in Figure \ref{fig:Classification_approach}, respectively.}
        \begin{tabular}{l  | c } 
            \toprule
            \textbf{Model} & \textbf{Accuracy (\%)} \\ 
            [0.5ex]
            \midrule
            ResNet-18 global (fw) & 76.08 \\
            ResNet-18 2S (fw) & 80.06 \\
            \midrule
            ResNet-18 global & 90.18 \\
            ResNet-18 2S & 90.49 \\
            \midrule
            %ResNet-50 2S (fw) & 80.675 \\
            %\midrule
            ResNet-50 global & 87.73 \\
            ResNet-50 2S & 88.98 \\
            \bottomrule
        \end{tabular}
    \label{tab:ml_model_performance}
\end{table}

\begin{figure}
\centering
\begin{minipage}{.48\textwidth}
  \centering
  \includegraphics[width=0.96\linewidth]{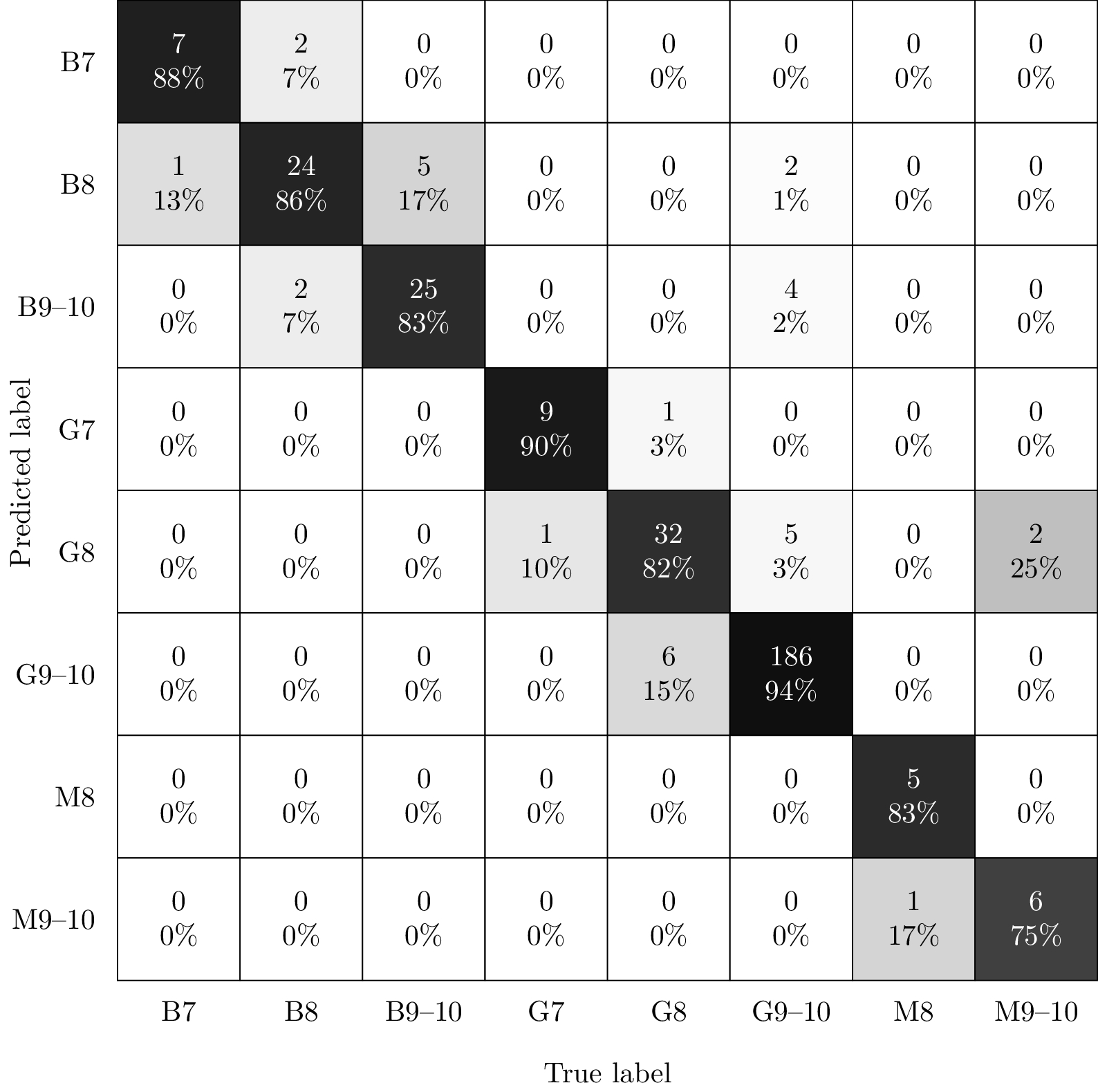}
  \captionof{figure}{The confusion matrix of the global ResNet-18 model with an overall accuracy of 90.18\% obtained on the test set.}
  \label{fig:CM_E2E}
\end{minipage}%
\hfill
\begin{minipage}{.48\textwidth}
  \centering
  \includegraphics[width=0.96\linewidth]{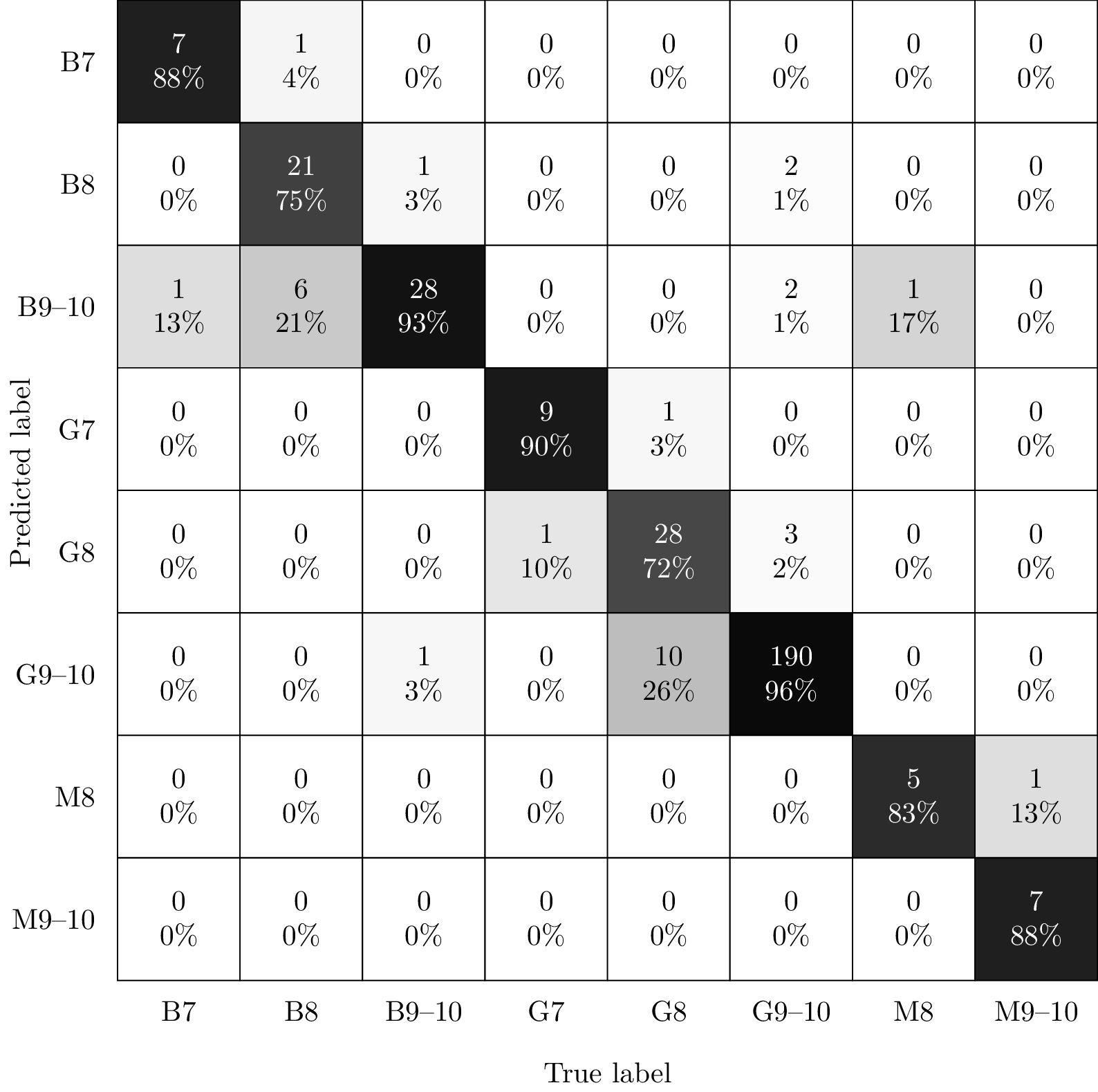}
  \captionof{figure}{The confusion matrix of the two-stage ResNet-18 model with an overall accuracy of 90.49\% obtained on the test set.}
  \label{fig:CM_2S}
\end{minipage}
\end{figure}

The combined accuracy for both tasks reaches up to 90.49\% using ResNet-18 as an architecture within a two-stage approach. On the dataset at hand, training the ResNet-18 architecture culminates in a better performance than ResNet-50, which in turn outperforms the case in which only the classifier portion of the ResNet-18 was optimized (fw). The two-stage approach marginally yet consistently outperforms the scenario in which a single model performs both classification tasks. It is evident that the merit of the two-stage approach depends on the architectural choice and the training strategy. For instance, the two-stage ResNet-18 (fw) with frozen feature extractor weights outperforms its global counterpart by roughly 4\%. In contrast, in both fully tuned examples, the performance improvement through the two-stage approach is less and amounts to 1.2\% and 0.3\% for the ResNet-50 and ResNet-18, respectively. The model's performances can be further dissected by Table \ref{tab:ml_model_performance_fine} and the confusion matrices provided in Figures \ref{fig:CM_E2E} and \ref{fig:CM_2S}. In the table, the model performances of each task-specific model contributing to the two-stage model are listed. 

\begin{table}[h]  
    \centering
    \caption{The prediction accuracies achieved on the test set for the different classification tasks. The values reported here are obtained by the individual elements of the two-stage (2S) models, see Figure \ref{fig:Classification_approach}b.}
        \begin{tabular}{l  | c | c | c | c } 
            \toprule
            & \textbf{Subtype} & \textbf{Grain size G} & \textbf{Grain size B} & \textbf{Grain size M} \\
            [0.5ex]
            \textbf{Model} & \textbf{Accuracy (\%)} & \textbf{Accuracy (\%)} & \textbf{Accuracy (\%)} & \textbf{Accuracy (\%)} \\ 
            [0.5ex]
            \midrule
            %%ResNet-50 E2E (fw) & 0.11 &  0.11 & 0.11 & 0.11 & 0.11 & 0.11 & 0.11\\
%          ResNet-18 2S (fw) & 90.58 & 88.21 & 71.21 & 64.29 \\
            %%ResNet-50 E2E & 0.11 & 0.11 & 0.11 & 0.11 & 0.11 & 0.11 & 0.11 & 0.11\\ 
            ResNet-18 2S& 96.35 & 93.90 & 86.36 & 92.86 \\
            %\midrule
            ResNet-50 2S& 96.96 & 93.09 & 81.82 & 92.86 \\
            %%ResNet-50 E2E & 0.11 & 0.11 & 0.11 & 0.11 & 0.11 & 0.11 & 0.11 & 0.11\\
            \bottomrule
        \end{tabular}
    \label{tab:ml_model_performance_fine}
\end{table}

From this table, it can be observed that the fully trained ResNet-18 variant improves on or matches the performance of the ResNet-50 variant throughout all grain size classification tasks. In contrast, the ResNet-50 performs slightly better at distinguishing the material subtypes. Moreover, it is apparent that the material subtype classification out of all tasks reaches the highest performance approaching 97\% accuracy. Thus, interestingly, the models manage to distinguish the martensitic, bainitic through-hardened, and martensitic through-hardened subtypes very well. This is reflected by the confusion matrix of the ResNet-18 2S model depicted in Figure \ref{fig:CM_2S} where it is illustrated that the remaining error cases predominantly arise from confusing the grain size categories (8 and 9--10) associated with small needle features in the B and G class. In terms of grain size classification, the three models dedicated to the three subtypes show some variation in their performance. On the larger G and B subsets, a performance of 93.9\% and 86.4\% is attained, respectively. In the confusion matrix in Figure \ref{fig:CM_2S}, it can be seen that the performance of the bainite grain size model falls short mainly because a notable portion of B instances was predicted as adjacent structure codes. The underlying reason for this will be further explored in the discussion section. Note, that the performance of the grain size models for martensitic grades (M) being seemingly unaffected by the architecture choice could be owed to the small size of the test set measuring only 14 images.

\section{Discussion}

\subsection{Round robin test and inter-rater reliability assessment}
A round-robin test was performed in which 14 metallographers trained on the needle length classification task participated in judging twenty micrographs. Those micrographs were selected from the test set based on the outcome of an intermediate DL model such that some correctly and some incorrectly classified micrographs were picked for each subtype and some grain sizes. In general, the images selected pose rather difficult instances since twelve out of twenty were misclassified by the intermediate deep learning model as well. This is in line with comments by the raters who deemed some images over-etched and thus difficult to classify. We opted for mostly identical classification conditions for the DL models and participants to facilitate comparability. The round-robin test entailed successively rating micrographs with respect to their subtype and grain size. In the case of subtype misclassification (e.g. `M' instead of `B'), the participants were presented with the consequential range of structure codes (M5--M10) rather than with the correct spectrum of structure codes (B5--B10) as answer options. Additionally, metallographers were able to provide information on which micrograph regions were involved in their grain size selection --- microstructural extrema or an average. Lastly, participants were able to provide free text comments for each image. All functionalities were integrated into a web application for straightforward participation, rating retrieval, and evaluation.\par

One important aspect to take note of is that the grain size label ('ground truth') for each image is determined by a single person. Therefore, the grain size label is afflicted by subjectivity as well. However, in contrast to the round-robin test participants, the original raters during day-to-day operation are provided with comprehensive background information about the specimen (alloy, heat treatment, storage, and contrasting) and flexibility during assessment (microscopy settings, observation of different specimen regions, and occasionally consulting further colleagues). While the decision about the reference grain size label is better informed than a single round-robin test participant, the reference label arguably does not represent a credible ground truth as it is still founded on the visual distinction of classes which are partly defined by nuanced changes in their image texture (see Figure \ref{fig:Overview_needle_morphologies}c--e) rather than quantitative measurements. In contrast, the subtype classification is deduced objectively from the heat treatment process parameters. However, the distinction between subtypes is not a task that the raters are commonly facing. Relying solely on images with micron bars rather than additional contextual processing information might incentivize the round-robin test participants to deviate from their daily classification habits and rely more on measuring needle length. Even the awareness of contributing to a round-robin test can result in a non-natural, to some degree disproportionate classification effort.
  
First, an overview of the round-robin test outcome for all micrograph instances is provided. Most errors were made during grain size assessment rather than subtype distinction. At first glance, this is surprising since the distinction between martensitic and bainitic microstructures is generally considered challenging and is not a task faced by the participating metallographers on a day-to-day basis. Specifically, out of 280 total predictions, the correct subtype was identified in 179 cases (63.9\%). Out of this subset, the correct grain size was determined in 56 instances (31.3\%). This corresponds to an overall accuracy of 20\%. When aggregating the grain sizes 9--10 and dropping the ones with gray background in Table \ref{tab:Needle_morphology_classes} (by discarding the corresponding instances with those predictions or labels leaving 187 predictions), the accuracy reaches 25.1\%. Considering the difficulty of the provided micrographs, the two successive classifications one of which is not commonly tackled during day-to-day operation, the noisy reference labels, and the vaguely defined decision thresholds for the mean or maximum microstructure criterion to be applied, the overall accuracy of 20--25\% is not striking. It was, however, unexpected to us that the subtype classification achieved a relatively high accuracy with 64\% under these circumstances, despite the DL models performing well at this task. This means that many metallographers picked up nuanced differences between these subtypes. Note that the 20 images were partitioned into 11, 5, and 4 for the G, B, and M subtypes, respectively. This skew might have simplified the subtype classification task due to the metallographer's awareness that the G subtype is dominating the data (see Supplementary Table 1) and their familiarity with the G material state. Indeed, the metallographers also performed better in the grain size predictions of the more common G and B subtypes and achieved  accuracies of 0.21 and 0.23 as opposed to 0.16 for the M minority class. The final global ResNet-18 model evaluated on the round-robin test images, which are rather difficult cases, achieves an overall accuracy of 70.59\% surpassing the 20--25\% achieved by human expert classification by a large margin. On a less difficult, more representative set, the classification accuracy of the raters would presumably increase.

\begin{figure}[htbp]
    \centering
    %\footnotesize
    \includegraphics[width=\textwidth]{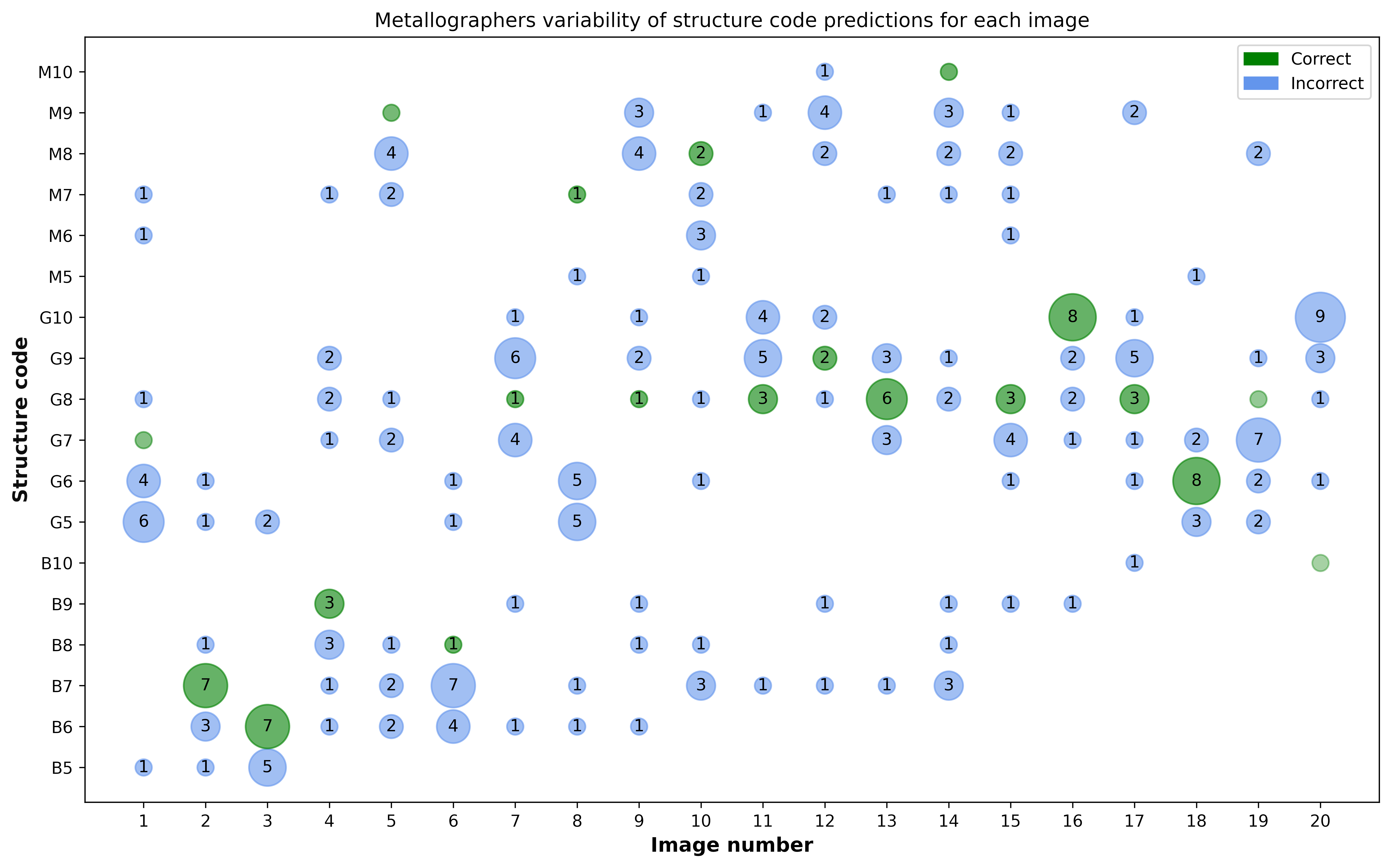} 
    \def\svgwidth{\textwidth}
    %\fontsize{12}{12}\selectfont
    %\input{mpnn_scheme.pdf_tex}
    \caption{A plot in which the circles indicate correct (green) and incorrect (blue) structure code predictions of the round-robin test participants for each image. The size of the circle and the contained number indicate the number of predictions for that specific class. For images where no grading is consistent with the reference label, an empty green circle is plotted to indicate the reference label.}
    \label{fig:Roundrobin}
\end{figure}

Figure \ref{fig:Roundrobin} shows the distribution of structure code predictions for each image. A majority vote in Figure \ref{fig:Roundrobin} coincides with the reference label for six image instances. Apart from that, there are eight image instances where a majority vote is adjacent to the reference label. In such cases, it is debatable whether the reference label or the majority vote is more trustworthy. Leaving aside the reference labels, in order to assess the inter-rater reliability, the overall Fleiss' kappa score was computed across all 14 participants. It amounts to 0.146, which is fairly low, even in view of the many classes, and thus a sign of elevated subjectivity and mediocre agreement between the participants. Only considering the material subtype distinction results in a higher Fleiss' kappa score of 0.330. When computing the overall Fleiss' kappa after reducing the possible classes, i.e. aggregating predictions and labels of grain size 9--10 and dropping image instances with grain sizes marked gray in Table \ref{tab:Needle_morphology_classes}, a value of 0.152 is obtained which is slightly higher than the 0.146. This is not surprising as the score depends on the cardinality of the categorical variables and the aggregation increases the accordance between the raters. Note that among the 14 participants, five deal with microstructures and grain size classification daily, while the others do less regularly (weekly to monthly). This affects the scatter of the round-robin test. When analyzing the overall Fleiss' kappa only for those five participants, the score increases from 0.152 to 0.318. This substantially higher accordance seems reasonable as these metallographers are not only more familiar with differences between the subtypes but also were more thoroughly trained to perform the grain size classification.

\subsection{Model performances}
As opposed to ratings by humans, the ResNet-18 and ResNet-50 models are deterministic and therefore repeatable as no stochastic elements such as dropout layers are used in their architecture. While at training-time stochastic online data augmentations are applied, at testing-time such operations were not utilized. The DL models supposedly learn an adequate mean representation and accurate decision boundaries from the noisy labels of many raters. An overall accuracy slightly exceeding 90\% is very satisfactory. This applies especially in view of the large data variance and the noisy, subjectivity-afflicted reference labels which limit the attainable performance on the test set. Training the complete ResNet models rather than relying on ImageNet weight initialization in the feature extractor allows for a better model specialization towards the downstream tasks which differ significantly from ImageNet classification. Thus, there is a pronounced performance boost for the completely optimized models as opposed to the ones where only the classification head was optimized, see Table \ref{tab:ml_model_performance}. The merit of the two-stage model was shown to be larger (4\% improvement) in the frozen weight scenario. This can be ascribed to the fact that tuning weights of a single fully-connected classification head in the global approach is presumably inappropriate for the present set of diverse tasks and classes. On the contrary, the performance of the fully trained models is likely most of all limited by annotation noise, thus overshadowing the influence of the two-stage approach.

The two-stage approach is motivated by the presumption that the morphology classification requires an emphasis on different features than distinguishing between martensite and bainite subtypes. It is anticipated that the latter relies on the distribution of retrained austenite and carbide phases, while needle morphology classification relies on length scales in the hierarchical substructure. Since the subtypes exhibit distinct microstructure length scales, decomposing the morphology classification further into three subtype-specialized grain size models was deemed promising. However, the results show that the two-stage model only marginally outperforms the global model and it is arguable whether it justifies the additional effort of training four distinct models. The two-stage approach facilitates picking the appropriate architectures for both tasks, depending on the representation power needed. In our case, a comparatively more expressive ResNet-50, and ResNet-18 seem to be beneficial for the subtype and grain size classification stages, respectively, as suggested by Table \ref{tab:ml_model_performance_fine}. When exploring how the decision-finding process in the two-stage process differs from the global model, it can be observed that the two-stage model, as expected, decomposes both decisions and relevant regions. Three Grad-CAM activation maps provided in Supplementary Figure 2 visualize this behavior. The same G class micrograph is passed to the global model and both stages of the two-stage model and Grad-CAM maps are extracted from the last layer of each ResNet-18 model. It seems that the global model considers regions of retained austenite, which are virtually exclusive to the G class, to perform the subtype distinction and simultaneously a few prominent needles to infer the grain size. In contrast, in the sequential approach, high activations occur solely at regions with high retained austenite concentration in the subtype stage and only at distinct needles at the grain size stage. It can be observed that the subtype model effectively takes more retained austenite regions into account to form its decision than the global model. This stronger activation might be caused by comparatively more specialized convolution filters and the increased feature density could potentially help to increase the model's confidence.

Whether and how tasks should be partitioned depends on the similarity of the tasks. For the challenge faced here, multi-task learning (MTL) is a very interesting and related paradigm. It utilizes hard or soft model parameter sharing together combined with distinct, task-optimized classification heads to address multiple tasks and optimize multiple objectives with a single model (\cite{ruder2017overview}). MTL can be especially beneficial if the tasks are related and rely on similar image features (\cite{caruana1997multitask}). This is presumably the case for the grain size distinction tasks across the three different subtypes. In fact, in the G and B subtypes, the needle length thresholds defining the grain sizes are virtually identical and deviate slightly for the M subtype. This could incentivize using a single jointly optimized backbone with three classification heads rather than three distinct subtype-specific grain size models prospectively. Moreover, in the current approach, only the data subsets of the respective subtype are used to train each of the three grain size classification models. This represents a strong restriction in terms of data quantity, especially for the M grain size model. Training the largest portion of the architecture with joint datasets, e.g. in the MTL setting, might alleviate this problem. The data subsets for the three grain size classification tasks are expected to have distinct labeling noise patterns. In such cases, parameter sharing in the architecture's backbone can result in learning more general representations where the data subset-dependent labeling noise is ignored (\cite{ruder2017overview}). MTL might incentivize the model to learn the relevance of needle length due to the additional evidence provided by the supervisory signal of different subtypes, despite their distinct needle morphology. Based on the earlier discussion on the decision-finding process in the two-stage model, it is rather unlikely that the subtype and grain size classification tasks rely on similar enough features to justify extensive parameter sharing. Nowadays, self-optimizing MTL models have been developed which optimize their architecture to train specific layer's parameters jointly across different tasks whenever it is beneficial to the overall performance, see \cite{misra2016cross, ruder2017sluice}.

\subsubsection{Material subtype classification}
\label{sec:dis_material_subtype_distintinction}
When examining the images, some distinctive features between the different subtypes and commonalities between instances of the same subtype can be identified. Specifically, the martensitic instances (M class) do not exhibit any retained austenite and carbide constituents (neither carbides nor retained austenite) and the overall needle structure appears comparatively coarse. In contrast, in the bainitic through-hardened (B class) instances, carbides always occur while virtually none have retained austenite. Moreover, the B class is characterized by a fairly heterogeneous needle length distribution. In some cases, a bimodal needle length distribution with two distinct regions occurs which is exclusive to the B class. The martensitic through-hardened material (G class) is typically more homogeneous than B and has an overall finer needle morphology. It can contain both, carbides and retained austenite, depending on the grain size. In finer structured material (8--10) carbides occur, while in coarser martensite (6--7) retained austenite is typical, see Figure \ref{fig:Overview_needle_morphologies}.  

\begin{figure}[htbp]
    \centering
    %\footnotesize
    \includegraphics[width=\textwidth]{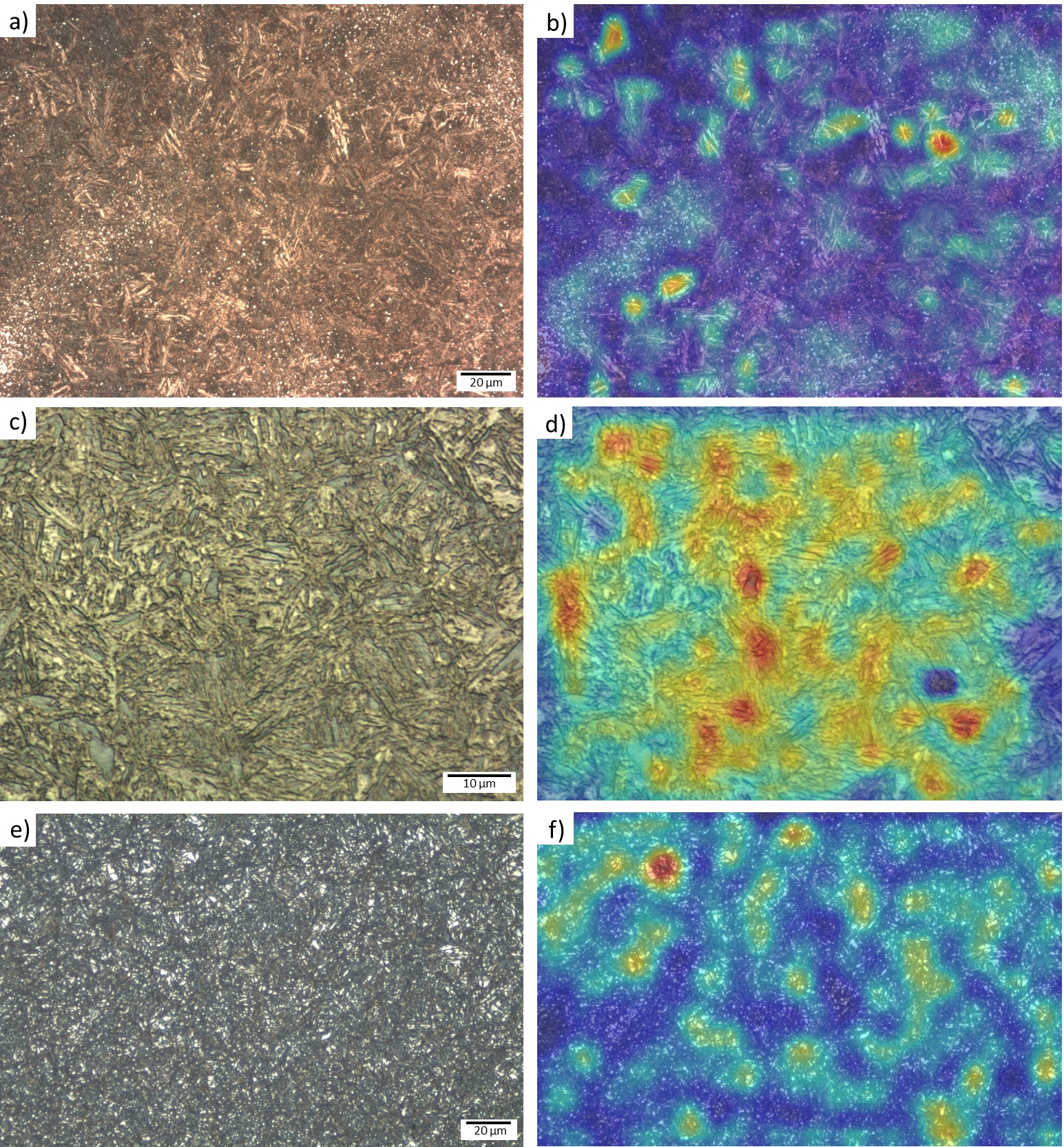} 
    \def\svgwidth{\textwidth}
    \caption{Steel micrographs showing different hierarchical microstructures (\textbf{A, C, E}) are input into the subtype classification model to extract the micrograph's Grad-CAM maps at the model's last layer (\textbf{B, D, F}). The Grad-CAM maps use `jet' colormaps. Thus, red, orange, and yellow colors highlight regions with increased activation.}
    \label{fig:subtype_distinction_analysis}
\end{figure}

It is interesting to explore whether the subtype model picks up on the same discriminative image features to achieve the accuracy of 97\% or whether it relies on other, maybe more subtle, aspects, such as slight differences in needle morphology induced by the varying quenching process. In order to investigate this, a few typical and atypical images of all subtypes are passed to the ResNet-18 subtype model, and the resulting Grad-CAM maps are illustrated in Figure \ref{fig:subtype_distinction_analysis}. 

In Figure \ref{fig:subtype_distinction_analysis}a and \ref{fig:subtype_distinction_analysis}b, a typical bainite micrograph and its activation map are shown. From the Grad-CAM map, it is evident that the subtype model concentrates on the coarser needle structures embedded within finer regions and on the clustered carbide regions. Both are typical characteristics of the B class. As opposed to this, in the M class micrograph and activation map depicted in Figure \ref{fig:subtype_distinction_analysis}c and \ref{fig:subtype_distinction_analysis}d, the attention of the model is spread more uniformly across the image. This is plausible as the C56E2 martensitic material seemingly exhibits a single-phase and relatively uniform microstructure with generally coarser grains. When assessing a through-hardened martensite (G class) micrograph in Figure \ref{fig:subtype_distinction_analysis}e and \ref{fig:subtype_distinction_analysis}f, it can be seen that the model focuses on retained austenite regions that are exclusive to the G class. An image pair in Supplementary Figures 3a and 3b shows a misclassification where the B class micrograph was confused for a martensitic through-hardened (G class) grade. This misclassification is probably owed to the fact that the microstructure is exceptionally homogeneous for a bainitic micrograph with no visible variation in needle size. Therefore, it indeed seems the predictions are only incorrect when the micrographs deviate from the trends introduced at the beginning of this section. While the features associated with those trends are discriminative for the range of heat treatments and alloys presented here, there are naturally other bainitic and martensitic grades that do not comply with the trends outlined in the first paragraph. This raises questions regarding model generalization to arbitrary bainitic/martensitic steels, e.g. in which retained austenite/martensite-austenite islands or pearlite constituents occur in a bainitic matrix.

\subsubsection{Needle morphology classification}
\label{sec:dis_needle_morphology_classification}
\begin{figure}[htbp]
    \centering
    %\footnotesize
    \includegraphics[width=\textwidth]{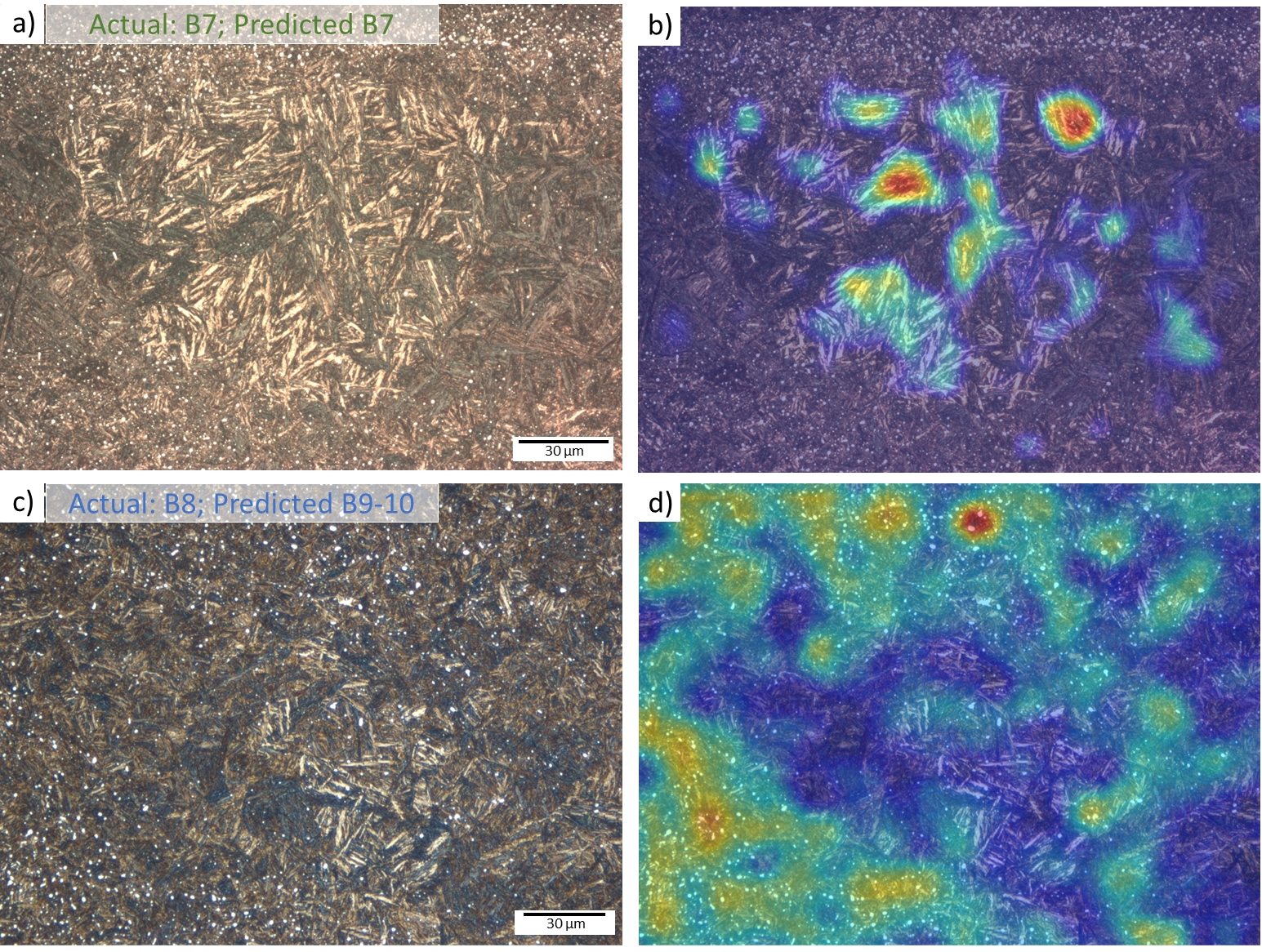} 
    \def\svgwidth{\textwidth}
    \caption{Bainitic steel micrographs showing heterogeneous microstructure (\textbf{A, B}) are input into the bainite grain size model to extract the micrograph's Grad-CAM maps at the model's last layer (\textbf{C, D}). The Grad-CAM maps use `jet' colormaps. Thus, red, orange, and yellow colors highlight regions with increased activation.}
    \label{fig:grainsize_distinction_analysis}
\end{figure}

The largest portion of the remaining errors can be ascribed to the grain size classification and annotation noise therein. Especially, the distinction between the finest grain sizes, G8/G9--10 and B8/B9--10, seems to cause misclassifications. In the bainitic materials, these misclassifications can be primarily attributed to heterogeneous microstructures. Figure \ref{fig:grainsize_distinction_analysis} shows two bainite micrographs that have been passed to the bainite grain size classification model along with the resulting Grad-CAM maps. Notably, the model takes the coarser regions into account to perform a correct prediction, see Figure \ref{fig:grainsize_distinction_analysis}a. This is a typical model behavior as the model learns to apply the maximum criterion (see Table \ref{tab:Needle_morphology_classes}). In a few cases, however, such as the one illustrated in Figure \ref{fig:grainsize_distinction_analysis}c--d, when the coarser regions occupy a small portion of the micrograph, the model tends to consider the finer regions to perform the prediction. While utilizing some kind of area threshold is the desired behavior, and it is promising that a lower area proportion of coarse regions leads to the model tending to the finer grain size prediction, the reference label was selected based on the coarse regions in this case. Instead of the eventually applied padding pre-processing approach, initially, we attempted to tile the raw images to fixed tile sizes and aggregate the tile's predictions. However, due to the heterogeneity of some micrographs in terms of needle length (see Figure \ref{fig:grainsize_distinction_analysis}), the tiling introduced further labeling noise and thus reduced the performance.

\subsection{Implications for quality control of microstructures}

While there are significant fluctuations in metallographic cross-section preparation and some variations in the alloying, there is also a notable dataset bias since the data is drawn from repeatable production processes. Specifically, the heat treatments culminate only in a small portion of possible martensitic and bainitic microstructural states. Indeed, this represents a simplification as the few distinct heat treatments result in fairly obvious discriminative features in the microstructure, see section \ref{sec:dis_material_subtype_distintinction}. Through interpretability analysis, models were shown to rely on these straightforward distinctive regions rather than learning nuanced correlations in the needle morphology. Thus, further data with more diverse martensite and bainite instances might be necessary to achieve better model generalizability by incentivizing the model to depend on a set of discriminative features for subtype distinction. When the dataset becomes vaster in terms of heat treatments and alloying, the distribution of retained austenite and carbide phases might not be a sufficient discriminative feature anymore. In such a case, adding a needle length objective as an additional task in an MTL setting could promote learning more nuanced differences in needle morphology between bainitic and martensitic states (\cite{ruder2017overview}).

Besides a high model accuracy, we consider three additional points particularly important when it comes to deploying such a model in a productive environment.

\begin{enumerate}
    \item Images that look similar to hardened bainitic or martensitic microstructures such as equivalent micrographs with some contained detrimental phases or a fully pearlitic micrograph should either be rejected from being classified by the model or a warning should be issued, that the image probably is not within the model's training distribution. This could help to prevent false classifications and raise confidence in the prediction's correctness as deep learning models otherwise were shown to exhibit mediocre generalization to out-of-distribution (OOD) samples (\cite{torralba2011unbiased}). In literature, depending on the degree of deviation and the available labels, the task of identifying abnormal images is distinguished into near/far OOD, anomaly, or novelty detection, see \cite{ruff2021unifying, bepari2017surface}. Typically, these tasks operate under the condition that a large quantity of OOD samples is unavailable which rules out training a binary OOD classifier. This condition is fulfilled in the present case as outliers are rare in production. However, subclass labels (e.g., steel subtypes) are available. Also, there is a necessity of detecting micrographs with subtle and local microstructural differences.  For these reasons, the task could be framed as near-OOD detection. Data-driven models in literature address OOD detection by measuring distances between image embeddings or by using reconstruction-based approaches  \cite{lee2017training,liang2017enhancing, bepari2017surface, ruff2021unifying}. While far-OOD detection has been successfully tackled for a range of domains, near-OOD detection still represents a major challenge, especially when nuanced local changes in unstructured images are concerned (such as the emergence of inclusions or some pearlite in microstructures).  Prospectively, such approaches could improve the model's robustness against arbitrary microstructures or metallographic artifacts. Images classified as OOD can then be inspected manually by metallographers.
    \item In order to continuously improve the model or to review individual model ratings by metallographers and data scientists, software solutions should facilitate providing process data. Dubious or interesting micrograph cases can then be collected to improve the model in the future. Aside from this, micrographs classified as in-distribution can be used for on-the-fly optimization within a semi-supervised learning framework. The process data should comprise saliency maps or other interpretability techniques. Supplementing information on the decision-making process is essential to establish trust in the proposed data-driven methodologies, especially when quality control is concerned.
    \item Nowadays, microscope systems are often connected to imaging software that assists users to adjust, acquire, and store images. Usually, additional functionality such as measuring or annotating image features is provided, which supports microstructural analysis and reporting. Integration of DL models into such software solutions will increase model acceptance. However, the deployment of trained models relies on microscopy software vendors providing appropriate software interfaces. This is a feature that only a few microscopy platforms offer yet will be indispensable going forward. 
\end{enumerate}

Treating the grain size estimation as a classification task might be beneficial to obtain a simple measure for repeated quality assessment throughout the supply chain. However, in this case, the categorization introduces a significant subjectivity. An alternative to the grain size assessment after ISO 643 could be to extract the needle length \textit{distribution} and derive physical metrics with more relevance towards target properties, such as probability distributions for rolling contact fatigue resistance. The micrographs at the magnifications required to judge the needle morphology cover a too small area to be representative of the whole material, especially for the more heterogeneous and coarser microstructures. Thus the current procedure requires the metallographer to identify the most critical region on the cross-section before image acquisition and needle assessment. This in itself introduces a subjectivity that is not considered in the round-robin test presented here.

\section{Conclusions}
A deep learning model was presented that categorizes hierarchically structured, quenched steels first with respect to their microstructure type and then their needle length. The model achieves satisfactory accuracy on both tasks and learns a mean representation of the data which is labeled by many metallographers, effectively reducing the impact of the significant labeling noise. Despite the thorough training of the metallographers, the task of manually assigning an ISO 643 grain size was demonstrated to be subjective in a round-robin test. Here, an objective and deterministic deep learning model can provide a remedy, especially if quality control throughout supply chains is concerned. The model's attention map is investigated by the presented interpretability analysis and consequences with respect to its generalization ability are discussed. For the deployment of the model, a robust out-of-distribution sample detection would be helpful since the model, at its current stage, is expected to not generalize across all possible martensitic and bainitic states. To achieve a more extensive generalization, data with fewer dataset biases, i.e. more diversity in terms of alloying and heat treatments than production data, should be supplemented.

\section{Materials and Methods}
\subsection{Dataset generation and statistics}
\label{sec:data_set_stats}
Although the basic process of metallographic microstructure assessment, namely preparing, etching, imaging, and evaluating, is standardized, it is not feasible to render the whole procedure entirely repeatable without automation --- especially in a production environment. Achieving this would require, amongst others, a reproducible storage period after polishing, fresh etchant for every specimen, etching times aligned to the millisecond, and the same brightness/contrast settings in the microscopes. Thus, even though all images were taken on upright metallographic microscopes after etching with Nital (2--3~\% alcoholic nitric acid solution) using the same imaging software, the image dataset features a significant variance. 

Predominantly, these variations result from using the individual magnifications, illumination, white balance, and luminosity settings that the different microscope/camera systems (ten different microscopes in this case) exhibit, as well as different etching times and etchant qualities applied by the respective users. Another major source of data scatter is that the images were captured at two plants that process different alloys. Nonetheless, these image dataset variations are still well within the natural limitations of the metallographic microstructure analysis process.

Last but not least, despite extensive training, every metallographer has a unique way of judging the images. After all, visual perception is very subjective, see \cite{anderson2011visual, panagiotaropoulos2014subjective}. Metallographers with decades of daily experience might not only comprehend the image textures differently than metallographers who rarely perform such tasks but also categorize them using a different approach. While very experienced metallographers will judge the needle coarseness alone by visual perception, inexperienced ones tend to apply a more quantitative approach in measuring needle lengths to correlate them with the grain size. Additionally, there are two possible rating systems applicable to judge the coarseness of the microstructure, as given in table \ref{tab:Needle_morphology_classes}. The first one is the mean criterion, where an image is rated according to the overall visual perception of the microstructure. This approach is pursued mainly if the depicted microstructure is homogeneous and fine-grained. In the case of a more inhomogeneous needle length distribution, and if the coarse-grained portion occupies a significant amount of the micrograph, it is possible to rate an image according to the maximum criterion. In that case, the coarse-grained regions in an image dictate the grain size label. This is motivated by the fact that ensembles of larger grains determine fatigue properties. When observing Figure \ref{fig:Overview_needle_morphologies}, it is apparent that the finer needle morphologies exhibit more similarity in their image texture due to their narrower decision boundaries.

\begin{figure}[htbp]
    \centering
    %\footnotesize
    \includegraphics[width=\textwidth]{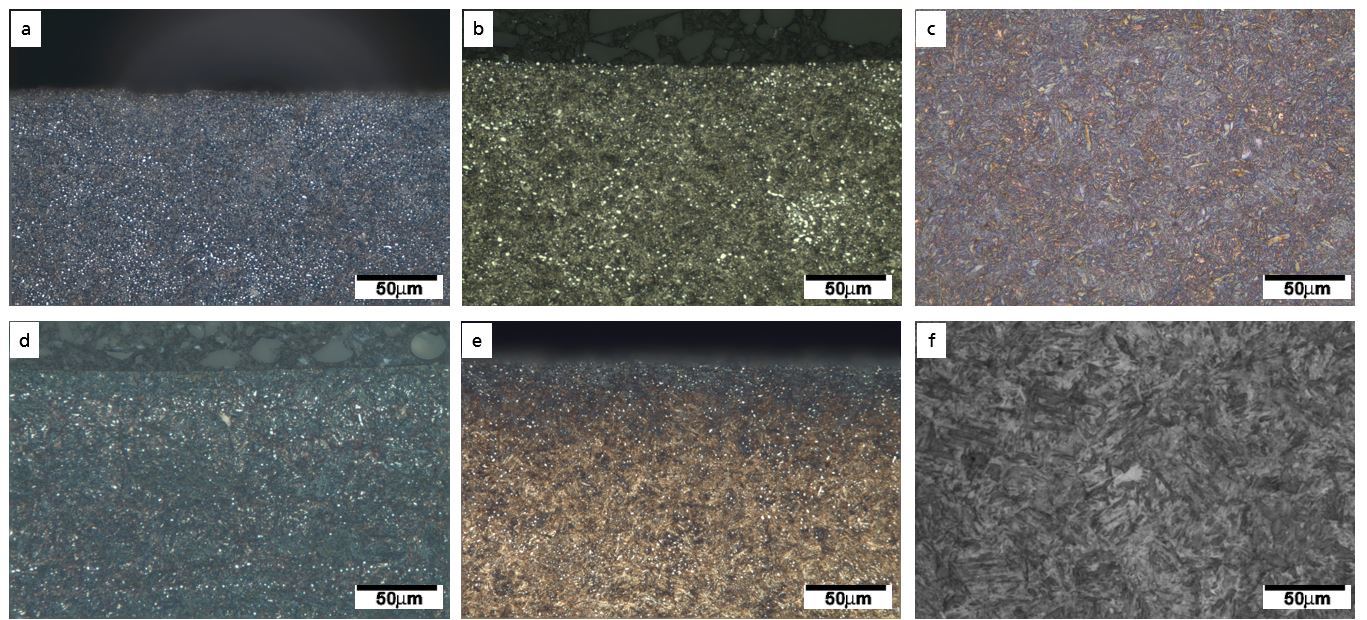} 
    \def\svgwidth{\textwidth}
    %\fontsize{12}{12}\selectfont
    %\input{mpnn_scheme.pdf_tex}
    \caption{A set of images depicting some martensitic and bainitic microstructures which underwent different hardening treatments with varying primary microstructure and distinct distributions of carbide and retained austenite constituents. The images are underlying the fluctuations described in section \ref{sec:data_set_stats}. Some images (\textbf{A--B} and \textbf{D--E}) cover regions outside the specimen at the top image border.}
    \label{fig:Overview_variance}
\end{figure}

In total, the dataset contains 1641 micrographs. A randomly sampled set of as-received images which is supposed to point out the contained variance is depicted in Figure \ref{fig:Overview_variance}. A summary with respect to the class distribution of the dataset is provided in Supplementary Table 1. There it is evident that the dataset is skewed towards the martensitic through-hardened G class (75.6\% of the overall data instances) and in particular to higher structure codes G9--10 (60.0\%). This is owed to the fact that this material is the predominant outcome of production. Aside from this, the bainitic through-hardened material ranks second (20.1\%). In contrast, the martensitic class M has a small share of the dataset (4.3\%). All three subtype datasets were split into train, validation, and test subsets using proportions of 64:16:20 in a stratified manner, i.e. ensuring that each subset contains a virtually identical distribution of structure codes. \par

There are further sources of data variance that are not considered in the data splits or later during imbalance correction. For instance, there is a pronounced imbalance in terms of alloys as depicted in Supplementary Figure 4. Roughly 68.7\% of the data represents 100Cr6 alloy. The remaining data is composed of 100CrMnSi6-4 (15.3\%) and otherwise distributed across twelve further 100Cr6 variants and the C56E2 alloy (M class). Some of the micro alloyed variants show up in the dataset with less than 10 instances and are hence clustered into a miscellaneous category in Supplementary Figure 4. Aside from this, the two magnifications with 500$\times{}$ and 1000$\times{}$ are applied depending on the appropriate field of view and spatial resolution for the microstructure at hand. Overall, the 500$\times{}$ images outweigh the higher magnification as they constitute approximately 82.4\% of the data. The magnification of $1000\times{}$ is thus most frequently employed for structure codes 9--10. Also, the fluctuation in terms of specimen preparation and contrasting conditions (etching) was not traced and are not given special attention during data preparation. While the utilized microscope setup for each micrograph is documented, no means of data adjustments, e.g. correction of optical distortion, were employed.

\subsection{Data pre-processing}

In Figure \ref{fig:Overview_variance}, some images were shown to feature regions outside the region of interest. Namely, the images typically contain micron bar annotations and often extend over the specimen borders. Therefore, either some defocused background or metallographic embedding resin regions are included which along with the micron bar can lead to spurious correlations. Since this can lead to the models learning non-causal relations, regions outside the region of interest were cropped in advance. In cases of an incoherent resin-sample-interface, so-called bleeding occurred occasionally where etchants or solvents creep out of the slit at the interface resulting in a visually altered surface region of the sample. This can be observed for instance in Supplementary Figure 1 within the green box annotation. Such regions were also removed in advance. \par

These modifications led to varying image resolutions. In order to cope with this, different strategies were tested in a preliminary ablation study. These strategies included tiling the images to a fixed size and aggregating the tile's predictions (1), resizing all images to the mean resolution of the dataset (2), and mirror padding/cropping to the most frequent resolution (3). As the latter turned out to perform best empirically, all images were transformed to 1994$\times{}$1994 resolution through mirror padding or minimal cropping. Similarly, to correct for the subtype and structure code class imbalances, see Supplementary Table 1, different imbalance correction methods were tested. These entailed instance-weighted cross-entropy and oversampling of the minority classes to balance out the data provided to the model. As in this case, oversampling the minority class performed slightly favorably, we carried on with it. \par    

Online data augmentations were applied during training before passing the color images to the models. The augmentation pipeline is composed of random rotations by arbitrary angles, random horizontal/vertical flips, slight random contrast adjustments, and Gaussian blurring. Subsequently, the images were normalized to the ImageNet mean and standard deviation. For the augmentations, the Albumentations package was used (\cite{buslaev2020albumentations}). 

\subsection{Model training and interpretability analysis}
All models were trained using ResNet-18 or ResNet-50 architectures. The models were initialized with ImageNet weights from torchvision (\cite{marcel2010torchvision}) and then either the full model was optimized with the same learning rate or the feature extractor portion of the classification network was frozen. In terms of learning rates, initially, 1E-4 or 1E-5 was used depending on the exact model. These learning rates were then modified by a StepLR scheduler. Cross-entropy was selected as the objective function. All submodels of the two-stage model were optimized individually and not trained end-to-end. When it comes to the two-stage approach, training the subtype model relied on the entire dataset, and training the individual structure code classifiers relied on the relevant data subsets. The models were trained for varying numbers of epochs until convergence, yet no overfitting was observed in the training and validation loss curves. The model performing best on the validation set was then used for the results reported here. 

In order to explore the tendencies of a model and understand the reasons for specific failure cases, model interpretability techniques can be helpful. In this work, we utilized a technique called Grad-CAM (\cite{selvaraju2016grad}). We applied it to explore the 
activation of the final convolution layer of the employed ResNet architectures. The technique provides heat maps where regions of pronounced activation in a specific image are highlighted. These heat maps are constructed by a weighted combination of all feature maps of that layer. The weights for each feature map correspond to the backpropagated gradients on which a global average pooling operation over width and height dimensions is applied.

\section*{Conflict of Interest Statement}

The authors declare that the research was conducted in the absence of any commercial or financial relationships that could be construed as a potential conflict of interest.

\section*{Author Contributions}
Conceptualization - AD, JM, and RN; Data curation - SP, DR, and RN; Formal Analysis - AD and SP; Funding acquisition - AD, JM, and RN; Investigation - AD and SP; Methodology - AD and SP; Project administration - AD and RN; Resources - AD and RN; Software - SP; Supervision - AD, JM, and RN; Validation - SP; Visualization - AD and SP; Writing (original draft) - AD, SP, and RN; Writing – review \& editing - AD, JM, RN, and SP 

\section*{Funding}
The work was tackled in a joint industry project between Fraunhofer IWM and Schaeffler Technologies AG \& Co. KG for which the latter party provided the funding.

\section*{Acknowledgments}
We want to express our gratitude to all metallographers of Schaeffler Technologies AG \& Co. KG who created the dataset and to all who participated in the round-robin test.

\bibliographystyle{unsrtnat}
\bibliography{bibliography}

\end{document}

\maketitle

\section{Supplementary Data}

\subsection{Details about the round-robin test}
When the round-robin test is evaluated per person, it becomes evident that the correct structure code outcomes range from a single one up to seven. At the subtype distinction, most raters still had 10 to 17 correct predictions with one outlier at 6 predictions. A few metallographers considered 2--4 micrographs over-etched impeding the classification. The participants made their choice regarding grain sizes predominantly (87\%) based on the overall impression rather than measuring individual coarse needles to then apply the maximum criterion.

\newpage
\section{Supplementary Tables and Figures}

%For more information on Supplementary Material and for details on the different file types accepted, please see \href{http://home.frontiersin.org/about/author-guidelines#SupplementaryMaterial}{the Supplementary Material section} of the Author Guidelines.

%Figures, tables, and images will be published under a Creative Commons CC-BY licence and permission must be obtained for use of copyrighted material from other sources (including re-published/adapted/modified/partial figures and images from the internet). It is the responsibility of the authors to acquire the licenses, to follow any citation instructions requested by third-party rights holders, and cover any supplementary charges.

%% Figures, tables, and images will be published under a Creative Commons CC-BY licence and permission must be obtained for use of copyrighted material from other sources (including re-published/adapted/modified/partial figures and images from the internet). It is the responsibility of the authors to acquire the licenses, to follow any citation instructions requested by third-party rights holders, and cover any supplementary charges.

\subsection{Tables}

\begin{table}[h]
        \centering
                \caption[Total number of images per class] {Total number of micrographs per class}
        \small
        \setlength\tabcolsep{3pt}
        \newcolumntype{P}[1]{>{\centering\arraybackslash}p{#1}}
        \begin{tabular}{P{0.20\textwidth}  P{0.20\textwidth} P{0.20\textwidth}  P{0.20\textwidth}}
            \toprule
\textbf{Subtype} & \textbf{Structure code} & \textbf{Number of micrographs} & \textbf{Total micrographs per subtype} \\% compensate for extrarowheight
           %\textbf{}  & \textbf{} & \textbf{}&\textbf{}& \textbf{} \\[-3pt]% compensate for extrarowheight
            \midrule
\multirow{4}{*}{Martensite 100Cr6 (G)}  & G6 & 11 & \multirow{4}{*}{1240} \\
& G7 & 48 & \\
& G8 & 196 & \\
 & G9-10 & 985 &  \\
 \midrule
\multirow{4}{*}{Martensite C56E2 (M)} & M6 & 0 & \multirow{4}{*}{71} \\
& M7 & 1 & \\
& M8 & 31 & \\
& M9-10 & 39 & \\
\midrule
\multirow{4}{*}{Bainite 100Cr6 (B)}& B6 & 4 & \multirow{4}{*}{330}\\
& B7 & 40 & \\
& B8 & 137 & \\
& B9-10 & 149 & \\
        \bottomrule
        \end{tabular}

        \label{tab:number of images per class}
\end{table}

\newpage
\subsection{Figures}

%%% There is no need for adding the file termination, as long as you indicate where the file is saved. In the examples below the files (logo1.eps and logos.eps) are in the Frontiers LaTeX folder
%%% If using *.tif files convert them to .jpg or .png
%%%  NB logo1.eps is required in the path in order to correctly compile front page header %%%

\begin{figure}[htbp]
\begin{center}
\includegraphics[width=0.8\textwidth]{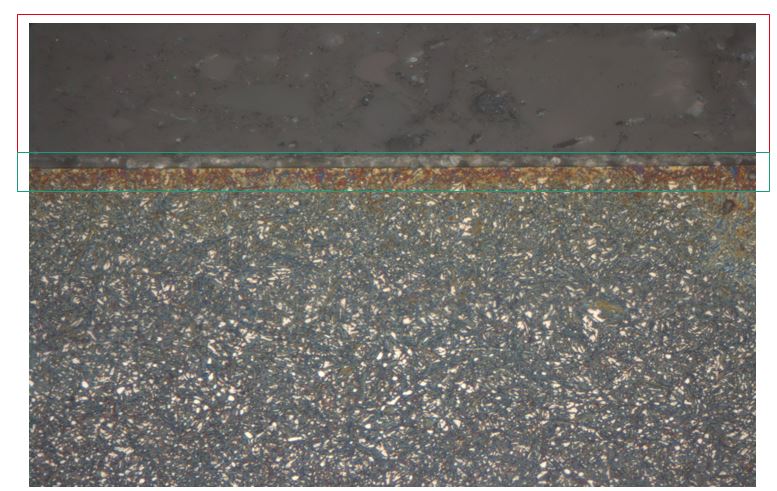}% This is a *.eps file
\end{center}
\caption{Regions in an as-received micrograph. The red box annotates the metallographic embedding resin out of the specimen and the green box the region where bleeding of etchants or solvents at the embedding interface altered the surface properties and thus the image texture.}\label{fig:1}
\end{figure}

\begin{figure}
    \includegraphics[width=.49\textwidth]{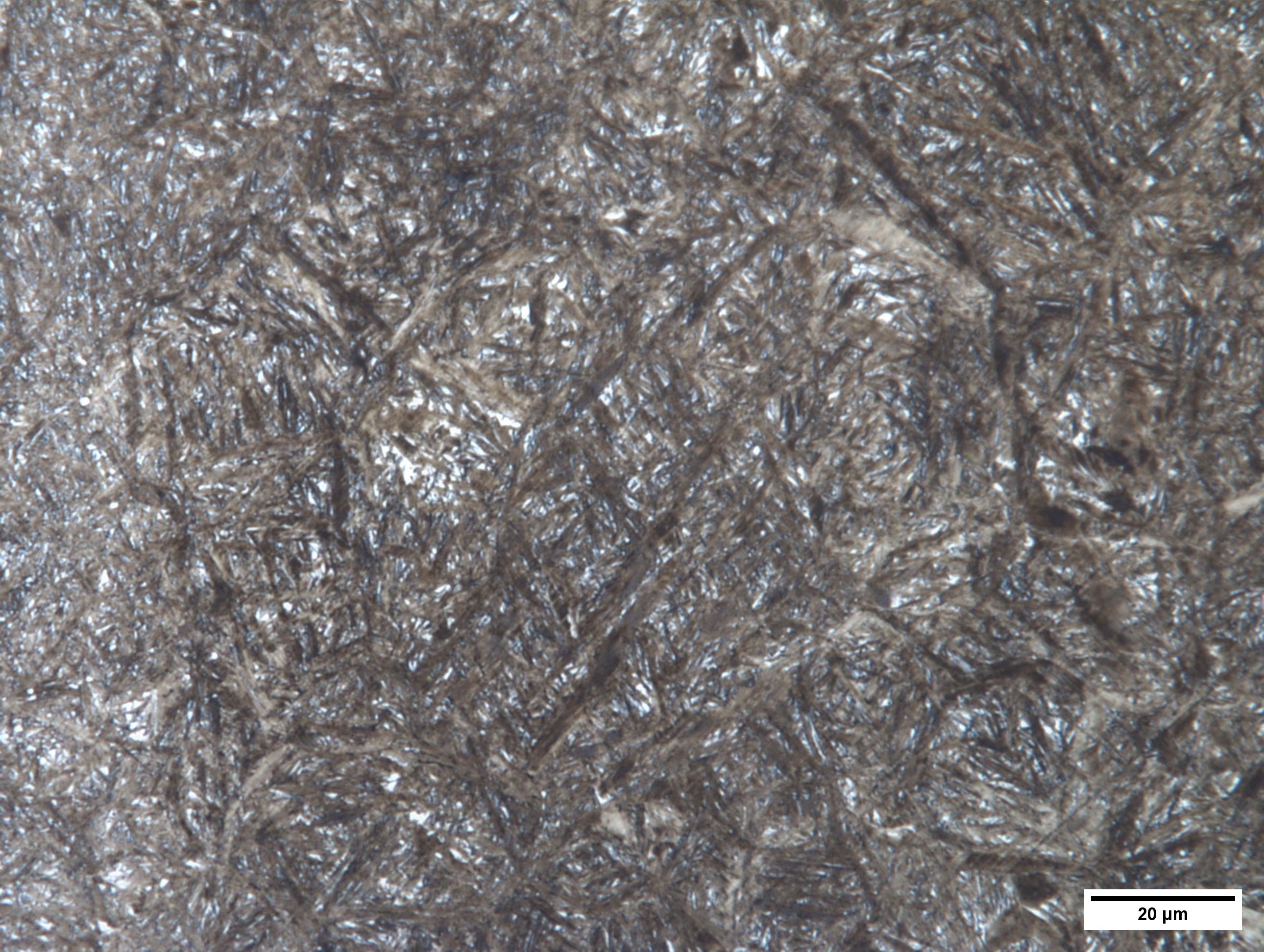}\hfill
    \includegraphics[width=.49\textwidth]{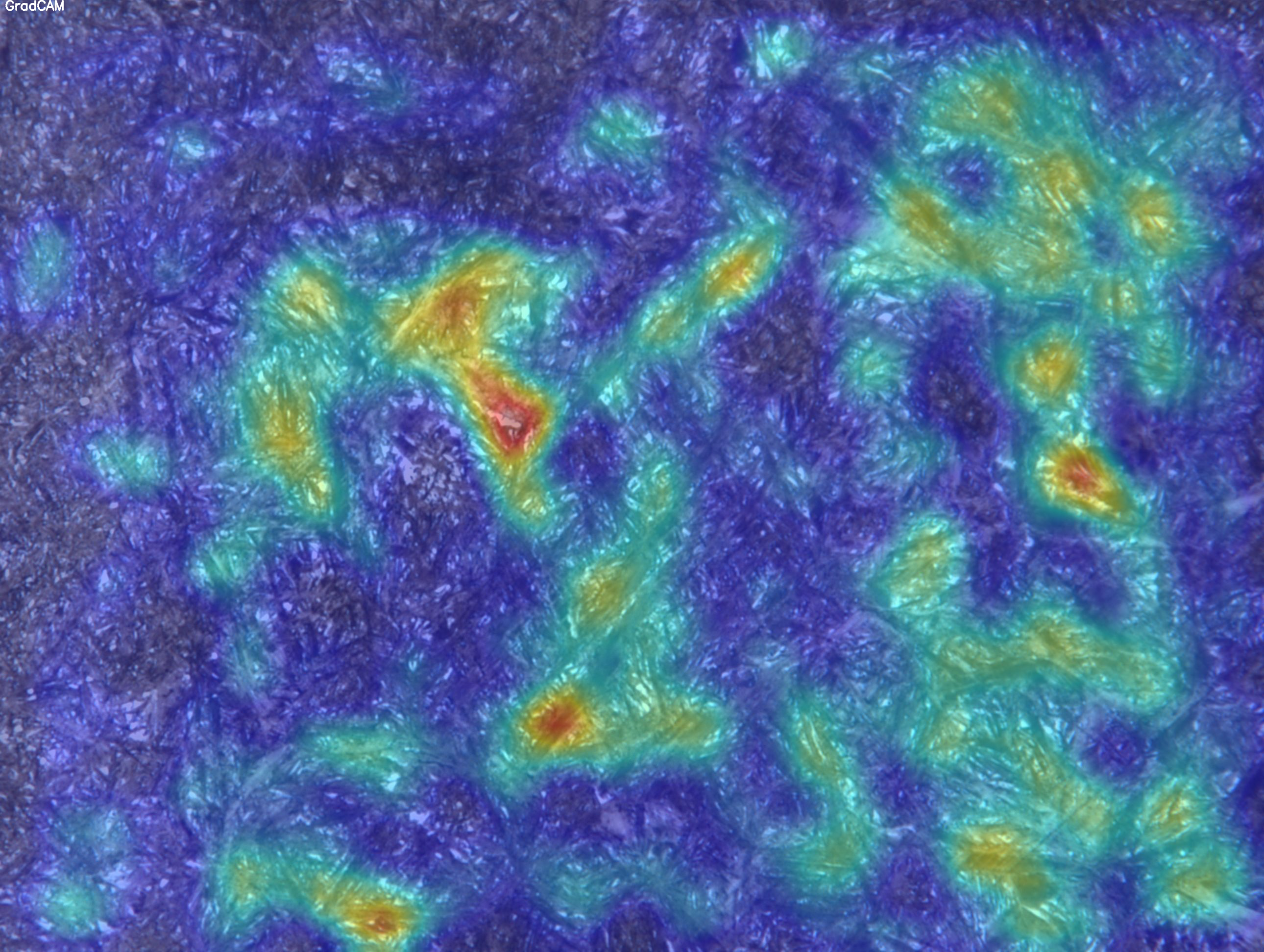}\hfill
    \\[\smallskipamount]
    \includegraphics[width=.49\textwidth]{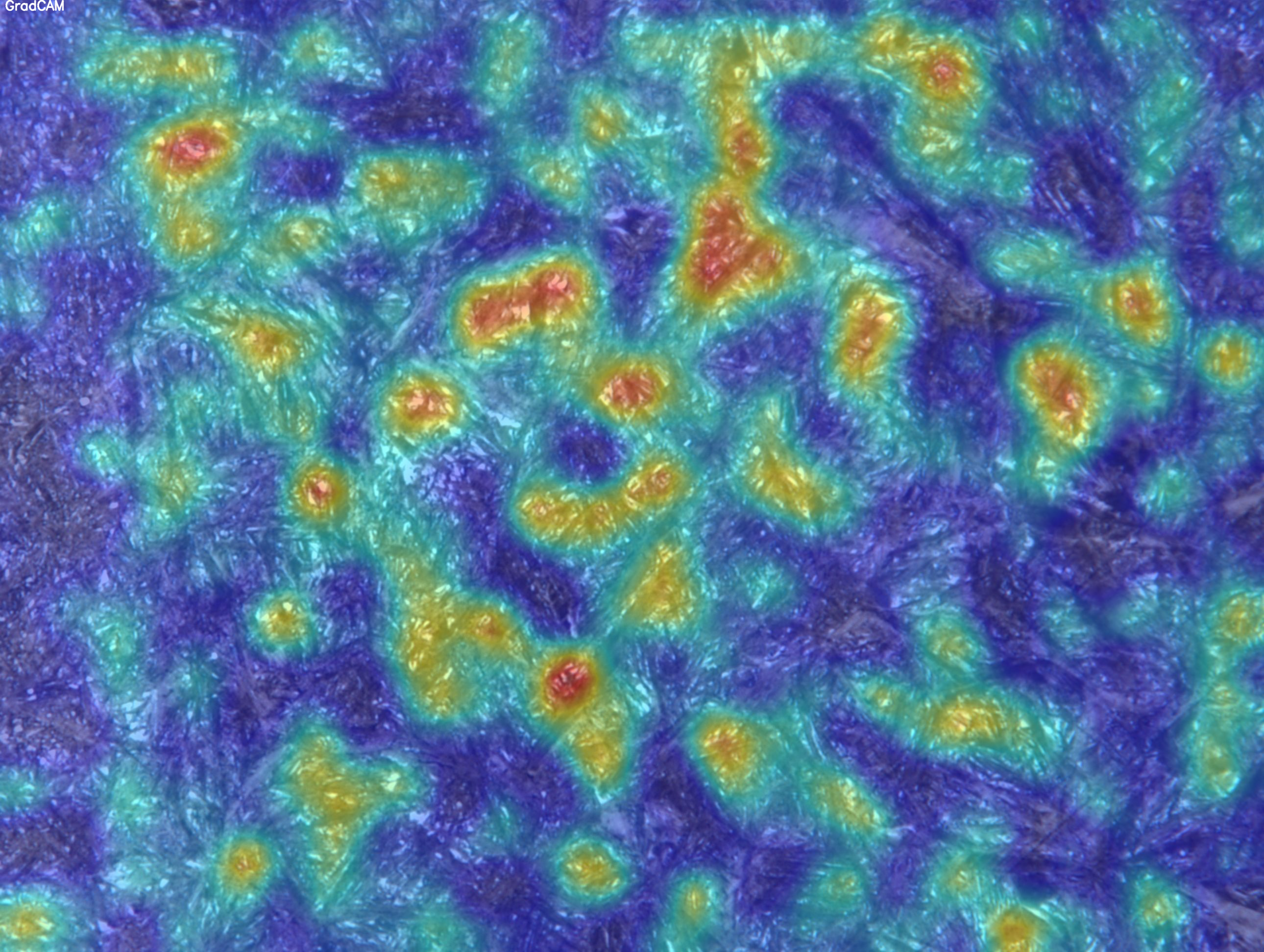}\hfill
    \includegraphics[width=.49\textwidth]{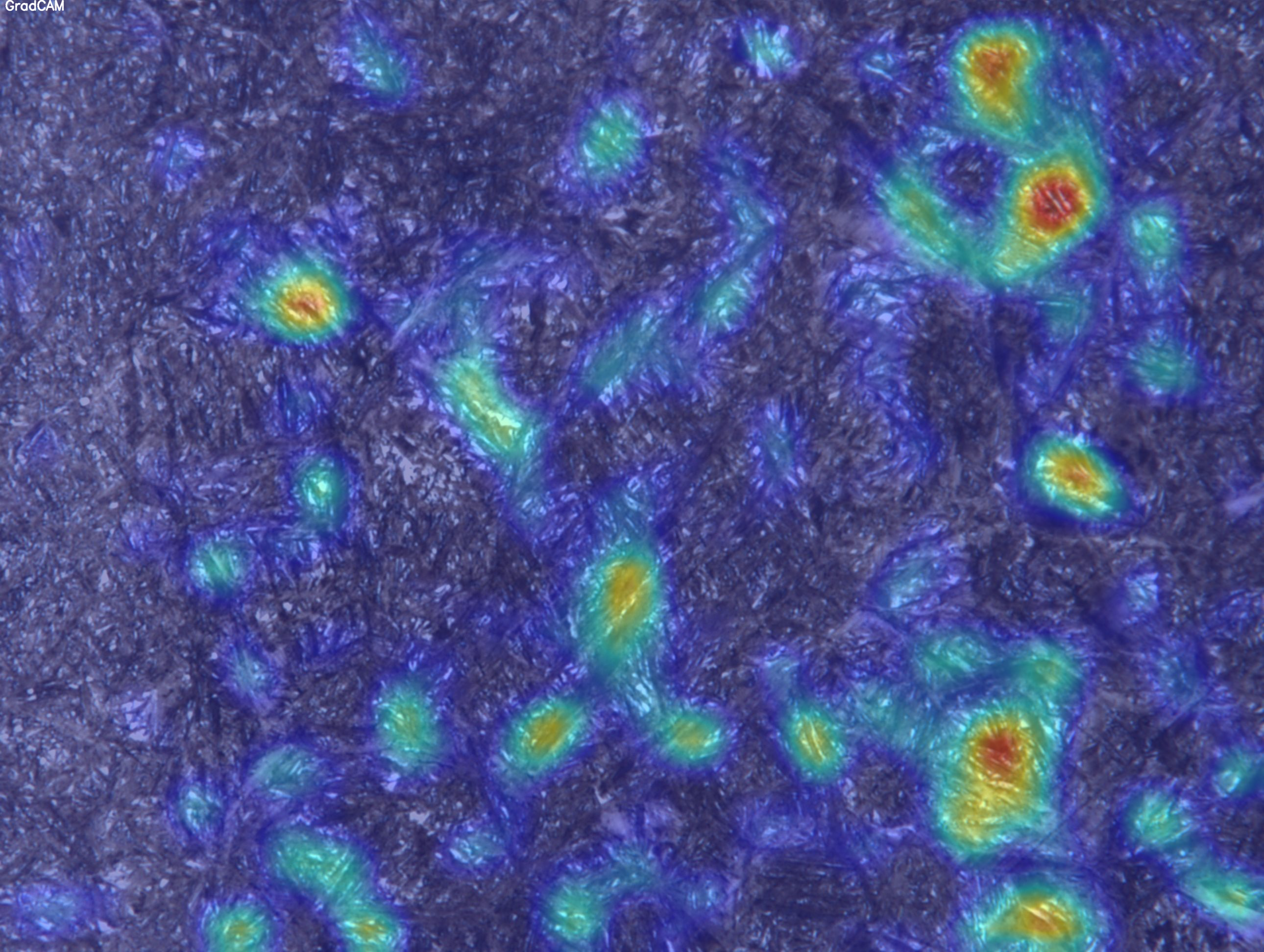}\hfill
    \caption{A martensitic steel micrograph (\textbf{top-left}) and its GradCAM maps extracted from the global model (\textbf{top-right}) as well as the subtype model (\textbf{bottom-left}) and grain size model (\textbf{bottom-right}) of the two-stage approach. Each GradCAM map is extracted from the final layer of a ResNet-18 model. The reference label of the micrograph in (\textbf{top-left}) is `G7'.}\label{fig:ETMWM_2013_0583_170_structurecode_gradcam}
\end{figure}

\begin{figure}
    \includegraphics[width=.49\textwidth]{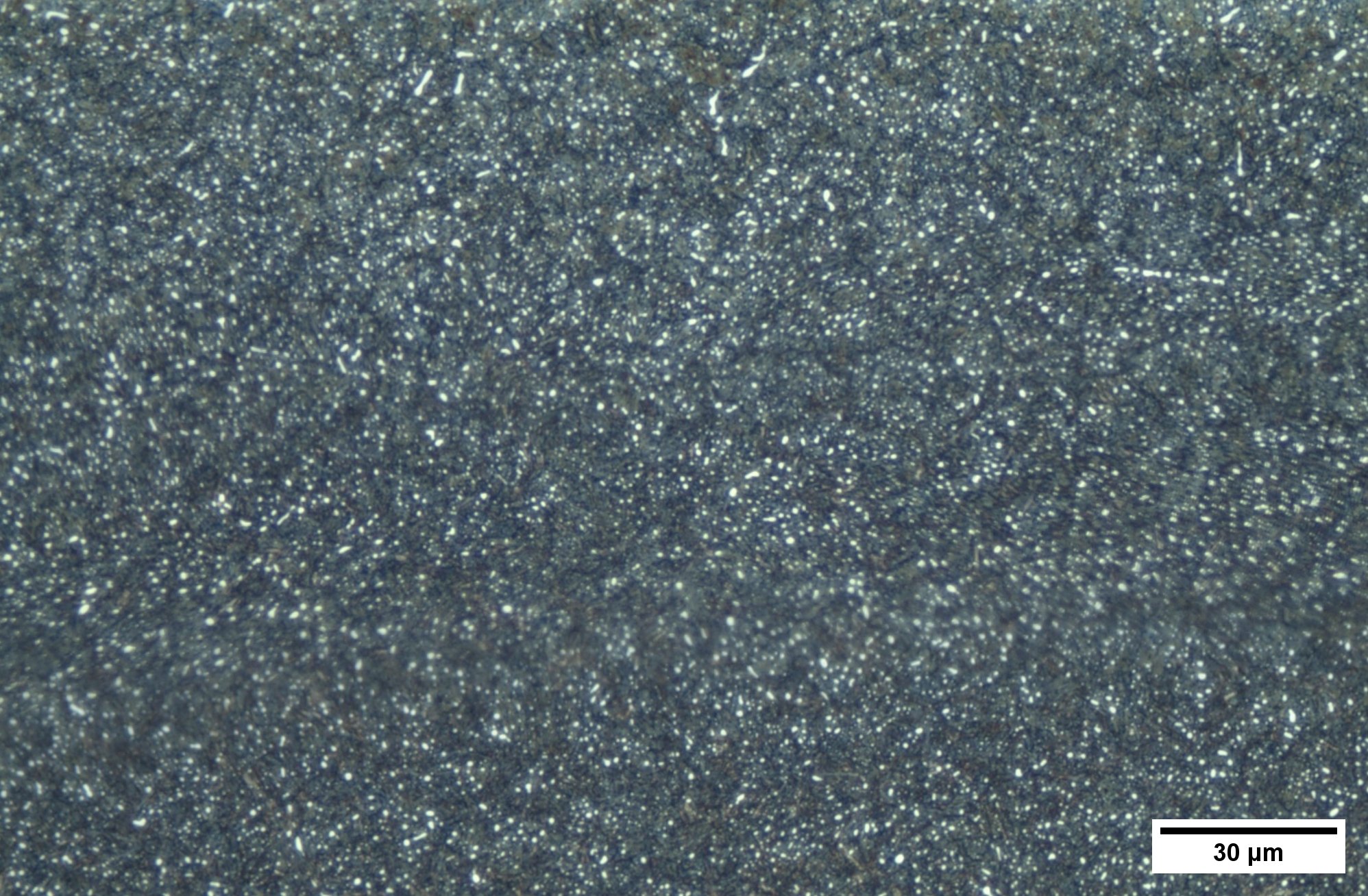}\hfill
    \includegraphics[width=.49\textwidth]{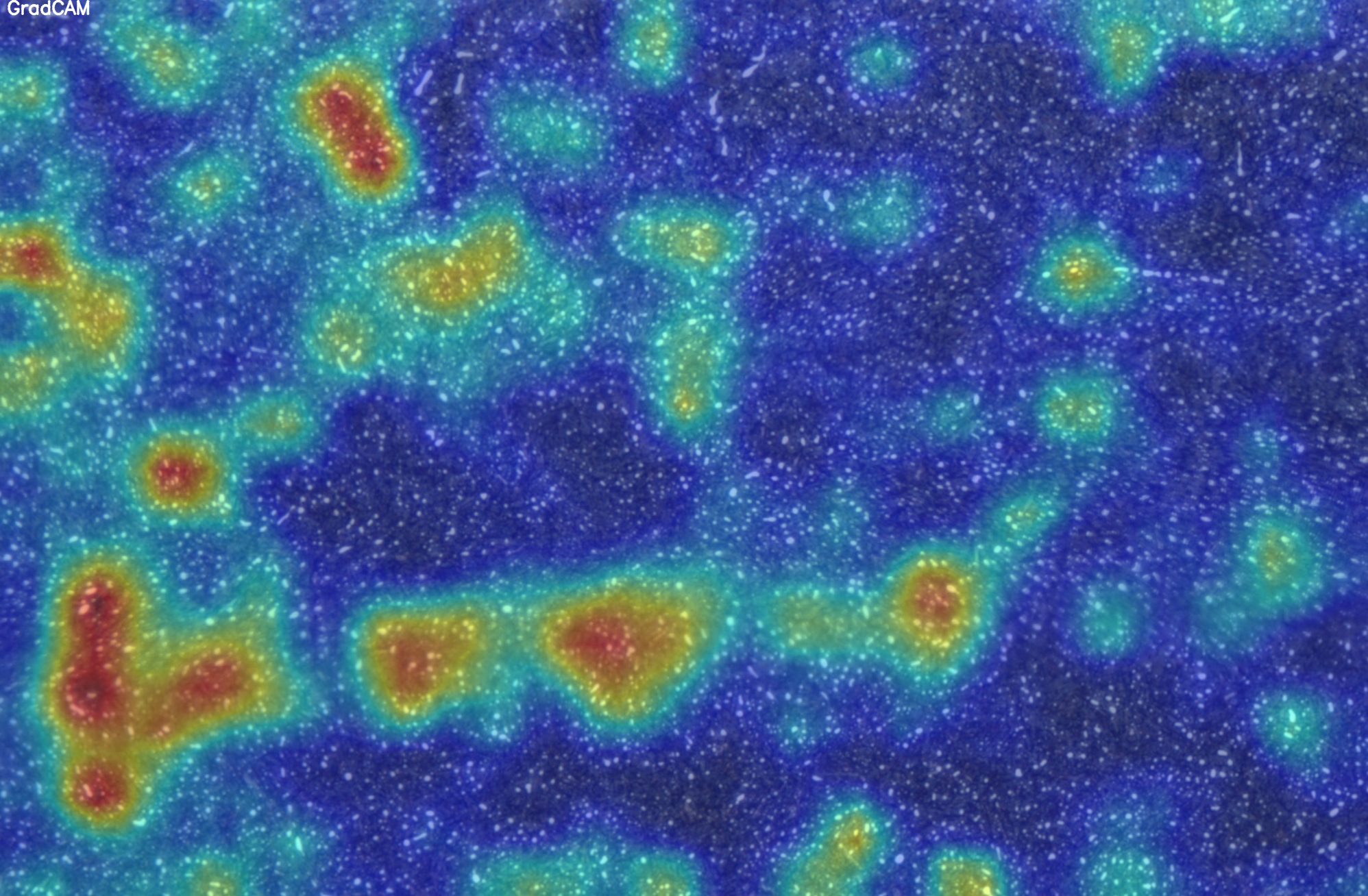}\hfill
    \caption{A bainitic steel micrograph (\textbf{left}) and its GradCAM map extracted from the subtype model (\textbf{right}) of the two-stage approach. The micrograph exhibits an image texture that rather resembles the ones present in the martensitic class G as it is significantly more homogeneous than other class B images. The GradCAM map is extracted from the final layer of a ResNet-18 model.}\label{fig:subtype_misclassification_analysis}
\end{figure}

\begin{figure}[htbp]
\begin{center}
\includegraphics[width=0.65\textwidth]{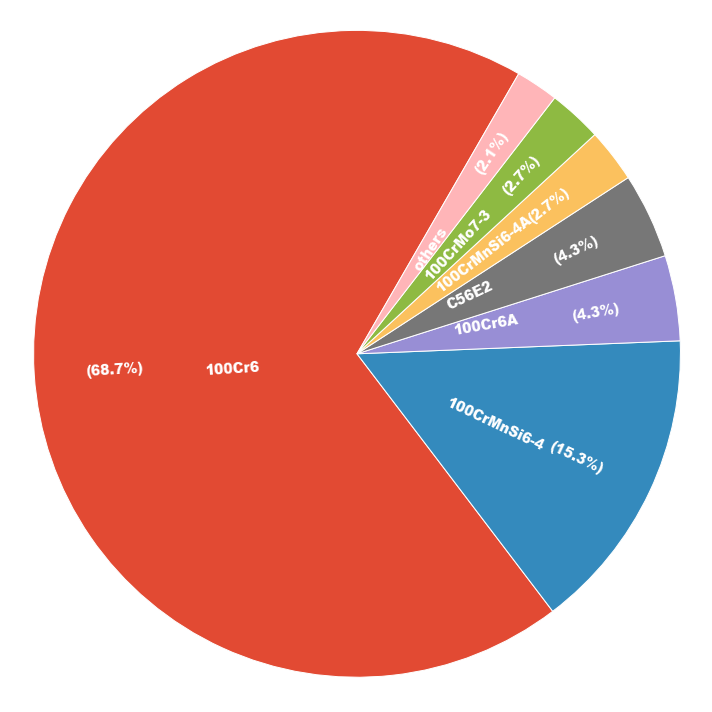}
\end{center}
\caption{Distribution of alloys in the dataset.}\label{fig:Alloy_distribution}
\end{figure}